\documentclass[journal=jacsat,manuscript=article]{achemso}

\usepackage{chemformula}
\usepackage[T1]{fontenc}
\usepackage{xcolor}
\definecolor{LightGray}{gray}{0.9}
\definecolor{tabgreen}{RGB}{44,160,44}
\definecolor{taborange}{RGB}{255,127,14}
\definecolor{tabblue}{RGB}{31,119,180}
\usepackage[separate-uncertainty=true,multi-part-units=single]{siunitx}
\DeclareSIUnit{\calorie}{cal}
\DeclareSIUnit{\bar}{bar}
\DeclareSIUnit{\kcal}{\kilo\calorie}
\DeclareSIUnit{\angstrom}{\text {Å}}

\author{Rolf David}
\email{rolf.david@ens.psl.eu}
\affiliation{PASTEUR, Département de Chimie, École Normale Supérieure, PSL University, Sorbonne University, CNRS, 75005 Paris}
\author{Miguel de la Puente}
\affiliation{PASTEUR, Département de Chimie, École Normale Supérieure, PSL University, Sorbonne University, CNRS, 75005 Paris}
\author{Axel Gomez}
\affiliation{PASTEUR, Département de Chimie, École Normale Supérieure, PSL University, Sorbonne University, CNRS, 75005 Paris}
\author{Olaia Anton}
\affiliation{PASTEUR, Département de Chimie, École Normale Supérieure, PSL University, Sorbonne University, CNRS, 75005 Paris}
\author{Guillaume Stirnemann}
\email{guillaume.stirnemann@ens.psl.eu}
\affiliation{PASTEUR, Département de Chimie, École Normale Supérieure, PSL University, Sorbonne University, CNRS, 75005 Paris}
\author{Damien Laage}
\email{damien.laage@ens.psl.eu}
\affiliation{PASTEUR, Département de Chimie, École Normale Supérieure, PSL University, Sorbonne University, CNRS, 75005 Paris}

\title{ArcaNN: automated enhanced sampling generation of training sets for chemically reactive machine learning interatomic potentials}

\begin{document}

\begin{abstract}
The emergence of artificial intelligence is profoundly impacting computational chemistry, particularly through machine-learning interatomic potentials (MLIPs).
Unlike traditional potential energy surface representations, MLIPs surpass the conventional computational scaling limitations by offering an effective combination of accuracy and efficiency for calculating atomic energies and forces to be used in molecular simulations.
These MLIPs have significantly enhanced molecular simulations across various applications, including large-scale simulations of materials, interfaces, chemical reactions, and beyond.
Despite these advances, the construction of training datasets — a critical component for the accuracy of MLIPs — has not received proportional attention, especially in the context of chemical reactivity, which depends on rare barrier-crossing events that are not easily included in the datasets.
Here we address this gap by introducing ArcaNN, a comprehensive framework designed for generating training datasets for reactive MLIPs.
ArcaNN employs a concurrent learning approach combined with advanced sampling techniques to ensure an accurate representation of high-energy geometries.
The framework integrates automated processes for iterative training, exploration, new configuration selection, and energy and force labeling, all while ensuring reproducibility and documentation.
We demonstrate ArcaNN's capabilities through two paradigm reactions: a nucleophilic substitution and a Diels-Alder reaction.
These examples showcase its effectiveness, the uniformly low error of the resulting MLIP everywhere along the chemical reaction coordinate, and its potential for broad applications in reactive molecular dynamics.
Finally, we provide guidelines for assessing the quality of MLIPs in reactive systems.
\end{abstract}

\section{Introduction}

The advent of artificial intelligence has revolutionized many fields of science, and machine learning has become an essential part of the scientific toolbox.
In computational chemistry, machine-learning interatomic potentials (MLIPs) now offer an attractive method that combines accuracy and efficiency for calculating atomic energies and forces, which are the computational bottleneck when running molecular simulations.
They have already led to remarkable successes, ranging from the simulation of very large-scale systems\cite{lu86PFLOPSDeep2021} to phase diagrams and transitions\cite{zhangPhaseDiagramDeep2021,piaggiPhaseEquilibriumWater2021,heStructuralPhaseTransitions2022}, metallic melts\cite{ryltsevDeepMachineLearning2022}, interfaces\cite{delapuenteAcidsEdgeWhy2022,wenWaterDissociationWater2023,delapuenteNeuralNetworkBasedSumFrequency2024,azomGrapheneOxideGraphene2024}, proteins in explicit solvent\cite{unkeBiomolecularDynamicsMachinelearned2024}, and chemical reactions\cite{zengExploringChemicalSpace2021,delapuenteAcidsEdgeWhy2022,youngReactionDynamicsDiels2022,devergneCombiningMachineLearning2022,benayadPrebioticChemicalReactivity2024,davidCompetingReactionMechanisms2024,gomezNeuralnetworkbasedMolecularDynamics2024,mondalModelingChemicalReactions2023,acharSilicoDemonstrationFast2023,zengMechanisticInsightWater2023,zhangIntramolecularWaterMediated2024}.

MLIPs provide a very high-dimensional fit of the potential energy surface (PES) of the system of interest, mapping the configuration space onto the potential energy.
Most of the computational cost is paid \textit{a priori} during the training of the model on a dataset that spans the range of important molecular structures\cite{bartokMachineLearningUnifies2017,chmielaExactMolecularDynamics2018,schranMachineLearningPotentials2021,keithCombiningMachineLearning2021,unkeMachineLearningForce2021,dingExploringChemicalReaction2024}.
The subsequent trajectory propagation then involves a much less expensive evaluation of forces with these potentials.
This therefore contrasts with other molecular dynamics methods which determine forces on-the-fly via costly calculations involving, \textit{e.g.}, electronic structure determinations, that need to be repeated for each configuration visited along the trajectory.

Over the years, a considerable effort has been devoted to the optimization of algorithms and network architectures, ranging from kernel-based methods\cite{mullerIntroductionKernelbasedLearning2001,
bartokGaussianApproximationPotentials2010,unkeMachineLearningForce2021,kaserNeuralNetworkPotentials2023} to high-dimensional neural networks and their many flavors\cite{behlerGeneralizedNeuralNetworkRepresentation2007,behlerAtomcenteredSymmetryFunctions2011,behlerRepresentingPotentialEnergy2014,behlerFourGenerationsHighDimensional2021,smithANI1ExtensibleNeural2017,devereuxExtendingApplicabilityANI2020,zhangDeepPotentialMolecular2018,zhangEndendSymmetryPreserving2018,schuttQuantumchemicalInsightsDeep2017,schuttSchNetContinuousfilterConvolutional2017,unkePhysNetNeuralNetwork2019,lubbersHierarchicalModelingMolecular2018,batznerE3equivariantGraphNeural2022,musaelianLearningLocalEquivariant2023,koFourthgenerationHighdimensionalNeural2021,zhangDeepPotentialModel2022,koAccurateFourthGenerationMachine2023}.
As a result of these recent developments, MLIPs now offer an attractive alternative to DFT-based\cite{tuckermanInitioMolecularDynamics2002,marxInitioMolecularDynamics2009} and reactive force field\cite{senftleReaxFFReactiveForcefield2016} molecular dynamics simulations.
While their computational cost is only moderately larger than that of classical force fields, they can be trained on high-level reference electronic structure calculations that provide much greater accuracy than empirical force fields.
Their efficiency is thus many orders of magnitude greater than that of DFT-based simulations.

However, while recent advances have considerably optimized the architecture of MLIPS and their descriptors, dataset construction -- another critical aspect affecting the quality of their energy and force predictions -- has not been as extensively explored.
Indeed, the training dataset should sample all typical configurations that will be visited during the simulation, while avoiding redundancies.

Different strategies have been adopted for the construction of the training dataset, depending on the type of processes to be simulated and on the available data.
In a first approach, the MLIP is trained only once, on a large collection of already available structures.
This is the case, for example, of the general-purpose potentials ANI\cite{smithANI1ExtensibleNeural2017,devereuxExtendingApplicabilityANI2020} and MACE\cite{batatiaFoundationModelAtomistic2024}, which are trained on a large dataset of chemically diverse organic molecules in their equilibrium geometry.
The resulting potential can then successfully describe the equilibrium fluctuations of a wide range of compounds in the gas phase.
However, larger geometric distortions that exceed the amplitude of thermal fluctuations are not included in the training dataset and are likely to be poorly described by the MLIP.

A type of active learning approach based on successive iterations, named concurrent learning\cite{zhangActiveLearningUniformly2019}, has thus been proposed.
Starting from an initial dataset, a first generation of MLIPs is simultaneously trained.
The latter are then used for explorations of the potential energy surface via unbiased molecular dynamics simulations, possibly under various temperature and pressure conditions.
In the configurations that are visited, the quality of the MLIP prediction is estimated by a query-by-committee approach,\cite{seungQueryCommittee1992} which measures the deviation among the predictions of the assembly of potentials that were trained on the same dataset (but with different random initializations).
Configurations in which the prediction uncertainty between the committee is large are then labeled with the reference calculation method and added to the training dataset for the next iteration of training and exploration.
This approach is, for example, successfully implemented in DP-GEN\cite{zhangDPGENConcurrentLearning2020} and expanded in ChecMatE\cite{guoChecMatEWorkflowPackage2023}.
We also note that recent uncertainty-aware and uncertainty-driven techniques have emerged as powerful tools for enhancing the accuracy and efficiency of MLIPs.\cite{schwalbe-kodaDifferentiableSamplingMolecular2021,xieUncertaintyawareMolecularDynamics2023,kulichenkoUncertaintydrivenDynamicsActive2023,vanderoordHyperactiveLearningDatadriven2023,zaverkinUncertaintybiasedMolecularDynamics2024}
By calculating the uncertainty of the MLIPs compared to the reference method, selecting configurations with high uncertainties, and possibly biasing the exploration of configurations toward poorly described regions, these approaches optimize the learning process, leading to more reliable and robust MLIPs, particularly in material science.
Other recent strategies, such as data distillation,\cite{anstineAIMNet2NeuralNetwork2024} have started to address the key component of constructing the training dataset.

However, a particular challenge is posed by chemically reactive systems, which require an accurate description of the energies and forces everywhere along the chemical reaction coordinate, including in the vicinity of high-energy transition states that are very rarely sampled spontaneously.
This difficulty is well known\cite{yangMachineLearningReactive2024}, and has started to be addressed by some first efforts.
A recent work\cite{zhangExploringFrontiersCondensedphase2024} has proposed a general-purpose reactive MLIP in condensed phases trained on a dataset including configurations collected over a wide range of temperature and pressure conditions.
Although this potential was shown to be successful for a number of chemical transformations, its exploration remains limited by the regions of the PES accessible via temperature and pressure changes, which implies that it is not adequate for chemical reactions with large energy barriers.
Another effort\cite{schreinerTransition1xDatasetBuilding2022} specifically sampled reaction pathways but was limited to reactions in the gas phase.
In a different approach, the training dataset can be enriched with configurations generated by enhanced sampling techniques\cite{yangUsingMetadynamicsBuild2022}, by performing random infinitesimal displacements\cite{youngTransferableActivelearningStrategy2021}, or by a combination of transition tube and normal mode sampling\cite{brezinaReducingCostNeural2023}. In a very recent work, a combination of uncertainty-driven dynamics and enhanced sampling was proposed to address reactivity at solid interfaces\cite{peregoDataefficientModelingCatalytic2024}.
All these strategies aim to explore the high-energy regions of the PES.
However, there is still a crucial lack of standardized procedures.
A set of uniform and consistent protocols would be needed to ensure that the training is easily reproducible, with proper bookkeeping of every file and parameter, and with a computational platform and workflow to support this.
Currently, each user must either manually or semi-automatically implement their own strategy, which becomes increasingly tedious for more complex systems, as constructing a reliable dataset typically involves many iterations.

Here, we address this major challenge for the efficient simulation of condensed phase chemical reactions.
We present ArcaNN, a comprehensive framework for generating training datasets for reactive MLIPs.
It combines a concurrent learning approach for the controlled convergence of the potential and a wide range of advanced sampling techniques for exploring the chemically relevant configurations, including high-energy geometries.
The exploration dynamics can be performed with either classical or quantum nuclear dynamics.
These successive steps are integrated into an automated approach that includes training, extended exploration, new configuration selection and associated energy and forces calculations at the reference level (labeling) steps, while keeping records so the procedure can be easily documented and replicated.

In the following, we first summarize the main steps of concurrent learning for MLIPs and describe the ArcaNN code, its architecture, and the different steps of the iterative training dataset generation.
We then illustrate its capabilities on a paradigm nucleophilic substitution reaction in solution.
We finally provide some concluding remarks about the applications and future developments of our code.

\section{Design of neural network interatomic potentials: overview}

\begin{figure}[h]
    \centering
    \includegraphics[width=8.3cm]{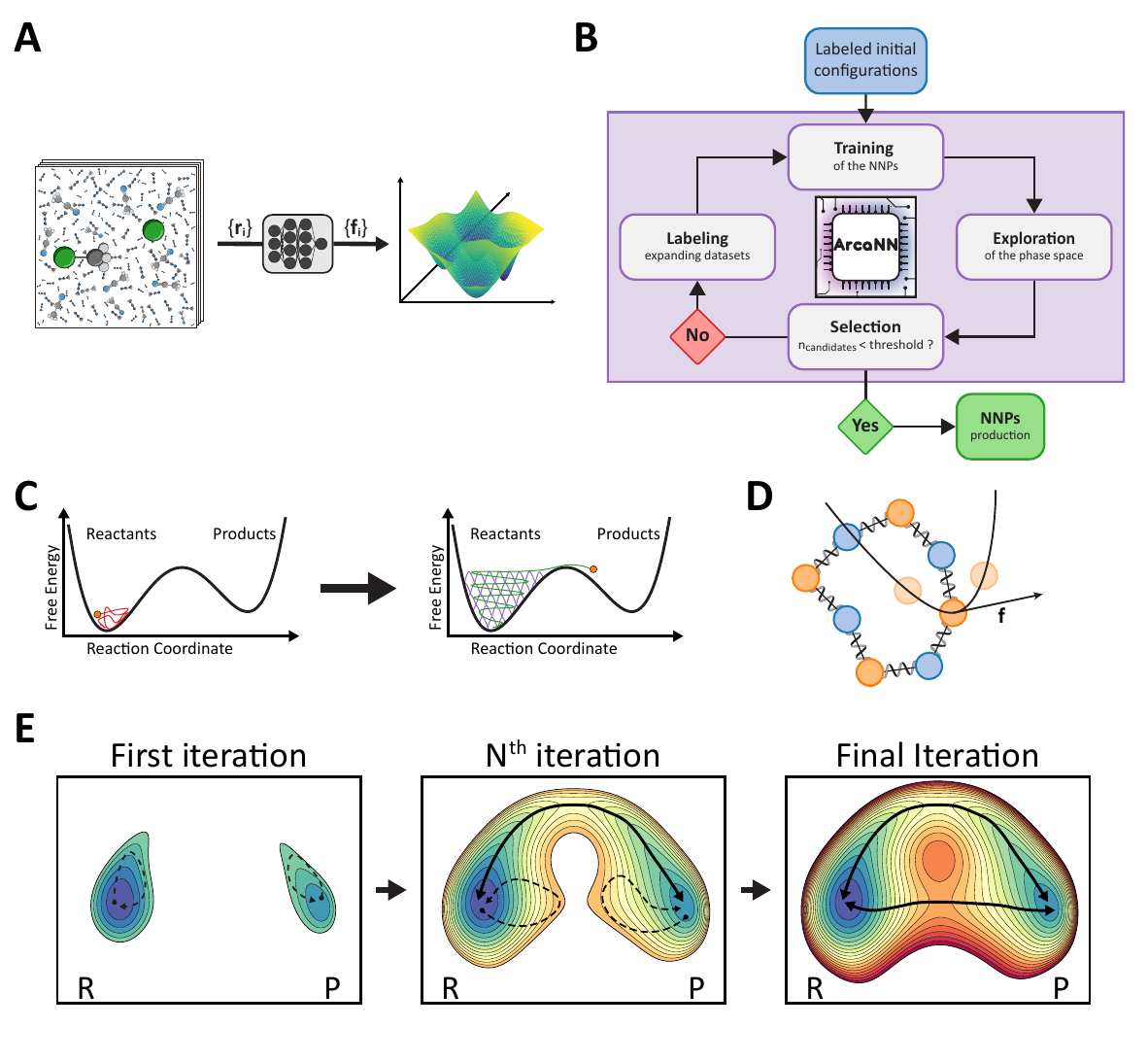}
    \caption{
        (A) Schematic representation of a neural network potential (NNP) that approximates the potential energy surface (PES) of a system.
        With a molecular structure as input, the NNP predicts the energy and forces of the system.
        (B) Schematic representation of the iterative training of NNPs using a concurrent learning loop.
        In the training, several NNPs are trained on a dataset of molecular configurations, each labeled with their corresponding energies and forces.
        During exploration, they are then used to run MD simulations and the selection phase assesses whether there are new candidates to be labeled to expand the datasets.
        The loop between training, exploration, and labeling can be repeated multiple times until there is no more candidates and the NNP is then deemed ready for production.
        In ArcaNN, the exploration phase is improved by the use of enhanced sampling techniques to explore the chemical phase space (C) and the possibility to perform path-integral MD simulations (D).
        This allows the iterative enrichment of the dataset, leading to a complete description of the chemical reactivity (E).
    }
    \label{fig_main:nnp_concurrent_learning_arcann}
\end{figure}

The objective of MLIPs, represented in Figure~\ref{fig_main:nnp_concurrent_learning_arcann}A, is to approximate the potential energy surface (PES) of a system.
For details regarding the different type of MLIPs architecture, the training and choice of descriptors for the atomic environment, we refer the reader to excellent reviews\cite{pinheiroChoosingRightMolecular2021,uhrinEyesDescriptorConstructing2021,raghunathanMolecularRepresentationsMachine2022,gokcanLearningMolecularPotentials2022,linDeterminationHyperparametersAtomic2023,tokitaHowTrainNeural2023,gomezNeuralNetworkPotentials2024}, of which we provide a brief overview below.

MLIPs have been developed based on different types of architectures, including artificial neural networks and kernel-based methods\cite{mullerIntroductionKernelbasedLearning2001,
bartokGaussianApproximationPotentials2010,unkeMachineLearningForce2021,kaserNeuralNetworkPotentials2023}.
A breakthrough in neural networks potentials (NNPs) came from the high-dimensional neural networks (HDNNs) introduced by Behler and Parrinello\cite{behlerGeneralizedNeuralNetworkRepresentation2007}.
The total energy of the system is decomposed into a sum of atomic contributions, which are assumed to exclusively depend on the local atomic environment encoded by a descriptor that satisfies the required PES invariances.
Two key advantages of this scheme and of this locality approximation are their computational efficiency and the possibility to extend these neural network models to arbitrarily large systems.
HDNNs with local descriptors based on a cutoff radius around each atom are used in several implementations, including BP-NNP\cite{behlerAtomcenteredSymmetryFunctions2011,behlerRepresentingPotentialEnergy2014,behlerFourGenerationsHighDimensional2021}, ANI\cite{smithANI1ExtensibleNeural2017,devereuxExtendingApplicabilityANI2020}, and DeePMD\cite{zhangDeepPotentialMolecular2018,zhangEndendSymmetryPreserving2018}.
Other MLIPs use the same atomic decomposition of the total energy but employ invariant message-passing neural networks (MPNNs)\cite{gilmerNeuralMessagePassing2017} for their descriptors; these implementations include, \textit{e.g.}, DTNN\cite{schuttQuantumchemicalInsightsDeep2017}, SchNet\cite{schuttSchNetContinuousfilterConvolutional2017}, PhysNet\cite{unkePhysNetNeuralNetwork2019}, and HIP-NN\cite{lubbersHierarchicalModelingMolecular2018}, which can access non-local information beyond the cutoff radius.
Recent improvements include the use of equivariant, atom-centered, message-passing neural networks, like NequiP\cite{batznerE3equivariantGraphNeural2022} and its evolution Allegro\cite{musaelianLearningLocalEquivariant2023}, which have been suggested to provide an improved accuracy compared to local approaches, and to remove the limitations on accessible length scales.
Finally, local models can also be extended by adding higher-order terms describing long-range effects and interactions\cite{koFourthgenerationHighdimensionalNeural2021,zhangDeepPotentialModel2022,
koAccurateFourthGenerationMachine2023,anstineMachineLearningInteratomic2023,chmielaAccurateGlobalMachine2023}.

NNPs are trained using a supervised learning approach, on an ensemble of molecular structures, each labeled with their corresponding energies and forces.
They usually demonstrate excellent accuracy in interpolating, \textit{i.e.}, predicting energies and forces for new configurations close to those seen during their training.
However, this accuracy drops dramatically when extrapolating to configurations not seen in the training, which is a key issue in machine learning models.
For molecular dynamics simulations, this implies that if the trajectory ventures outside of the training dataset region, the NNPs will typically lead to unphysically large forces that abruptly terminate the simulation.

This issue could be addressed by identifying all relevant configurations \textit{a priori}, for example, from an extensive sampling with a long simulation.
However, this requires being able to calculate the energies and forces during this long trajectory and necessitates, for example, \textit{ab initio} molecular dynamics (aiMD).
This solution is not practical since sampling with aiMD is computationally demanding, especially when the configurational space to be mapped is large.
In addition, propagating long trajectories with good accuracy for the force calculations is precisely the objective of NNP-based simulations.

To address this situation where the volume of unlabeled data can be large but the cost of labeling is high, an iterative construction of the training dataset inspired by the concept of active learning\cite{settlesActiveLearning2012} was proposed to navigate through the data, gather feedback, and proactively seek labels for data points that are marked as requiring further attention.
This concurrent learning approach\cite{zhangActiveLearningUniformly2019,zhangDPGENConcurrentLearning2020}, illustrated in Figure~\ref{fig_main:nnp_concurrent_learning_arcann}B, involves three main steps: exploration, labeling, and training.
These steps are repeated until convergence, which can be estimated using various descriptors and criteria.

However, exploration trajectories are usually propagated without any bias in the configurational space, and, as a consequence, chemical reactions with a free-energy barrier exceeding a few times the thermal energy do not spontaneously occur on the timescale of these simulations.
An additional limitation is that during a reactive trajectory, the time spent in the transition state region is very limited.
This unbalanced sampling therefore contrasts with the objective of a uniform sampling along the reaction coordinate to ensure that the error is low everywhere along the reaction path.
Another limitation is that chemical reactions are rare events, and a given reactive trajectory between reactant and product regions is often short-lived (on the picosecond timescale).
Finally, another difficulty is that for systems where several reaction pathways are in competition\cite{benayadPrebioticChemicalReactivity2024,davidCompetingReactionMechanisms2024}, we would like to sample all pathways and not only the minimal free energy one.

In order to better sample high free-energy barriers, enhanced sampling simulations are necessary.
Examples include, but are not limited to, umbrella sampling\cite{torrieNonphysicalSamplingDistributions1977}, metadynamics\cite{laioEscapingFreeenergyMinima2002} and its variants\cite{barducciWellTemperedMetadynamicsSmoothly2008,invernizziRethinkingMetadynamicsBias2020}, which have already been successfully applied in the context of data generation for NNPs\cite{yangUsingMetadynamicsBuild2022,devergneCombiningMachineLearning2022,delapuenteAcidsEdgeWhy2022,delapuenteHowAcidityWater2023,benayadPrebioticChemicalReactivity2024,delapuenteNeuralNetworkBasedSumFrequency2024,davidCompetingReactionMechanisms2024,zhangIntramolecularWaterMediated2024}.
Generally, these require identifying a set of collective variables (CVs) to bias the exploration trajectories, or setting up multiple enhanced sampling simulations covering numerous CVs to ensure that the reaction pathway is sampled adequately.

An important limitation in the current state of the art is therefore that users must either resort to a nano-reactor approach\cite{zhangExploringFrontiersCondensedphase2024}, which sacrifices control over specific reactivity and pathways, or they must manually set up numerous enhanced sampling simulations, which are both tedious and time-consuming.
This is the limitation addressed by ArcaNN.
It provides a comprehensive, flexible and automated workflow to generate datasets to train reactive NNPs while recording all the steps leading to the construction of the datasets, which can thus be easily shared and reproduced, a step towards meeting the FAIR principles\cite{wilkinsonFAIRGuidingPrinciples2016} for research data.

\section{Streamlining the construction of a reactive dataset with ArcaNN}

\subsection{Concept}

ArcaNN is a comprehensive framework, interfaced with other neural network, molecular simulation, and quantum calculation software for training NNPs, propagating trajectories, and labeling new configurations.
ArcaNN allows the sampling of the chemical phase space of a given reaction (encompassing reactants, products, intermediates, and transition states with the solvent treated in a reactive manner) to adequately and efficiently build a dataset that can be used to train NNPs.

The workflow combines the concurrent learning approach with enhanced sampling techniques, as shown in Figure \ref{fig_main:nnp_concurrent_learning_arcann}C.
Starting from an easily generated dataset of structures in the reactant and product regions, ArcaNN supervises the simulation of  either classical or path-integral swarms of short biased dynamics.
The dataset is progressively enriched with representative structures along the reaction pathways, on which generations of NNPs are iteratively trained and used for sucessive rounds of explorations.
This approach not only makes more efficient use of computational resources compared to an equivalent biased initial \textit{ab initio} trajectory but also provides a greater number of uncorrelated samples, leading to more accurate NNPs.

\subsection{Overview of the code and definitions}

ArcaNN is a Python 3 package designed in a modular fashion to facilitate its extension, modification, and the integration of new features, such as interfacing with new software or types of MLIPs.
The current version of ArcaNN is interfaced with the following programs:

\begin{itemize}
    \item CP2K\cite{kuhneCP2KElectronicStructure2020} for labeling;
    \item DeepMD-kit\cite{wangDeePMDkitDeepLearning2018,zengDeePMDkitV2Software2023} for training the NNPs;
    \item LAMMPS\cite{thompsonLAMMPSFlexibleSimulation2022} or i-PI\cite{kapilIPI20Universal2019} for exploration using the DeePMD NNPs, both in combination with Plumed\cite{tribelloPLUMEDNewFeathers2014} for enhanced sampling.
\end{itemize}

ArcaNN maintains a clear and easily readable record of the workflow.
This framework offers great flexibility at each workflow step, including the full range of quantum chemistry methods available in CP2K and the diverse enhanced sampling techniques and CV definitions offered by Plumed.
Users can also choose to explore any number of \textbf{systems}.
As detailed below, these correspond to a combination of MD parameters, thermodynamic conditions, and chemical compositions.

ArcaNN is specifically designed for High-Performance Computing (HPC) clusters with CPU and GPU resources, exploiting them in an embarrassingly parallel fashion.
It utilizes VMD\cite{humphreyVMDVisualMolecular1996} for trajectory manipulation in DCD format and Atomsk\cite{hirelAtomskToolManipulating2015} for converting LAMMPS data files to and from the XYZ format.

From the initial datasets and user-provided files, ArcaNN oversees the creation of necessary files and folders for the interfaced programs and submission scripts for HPC resources.
It manages the training of NNPs, the exploration of phase space, and the labeling of configurations, and it iterates these \textbf{steps}  until the NNPs accurately describe the targeted reactivity of a given system.
While requiring minimal intervention, ArcaNN gives users full control over the iterative process through a series of \textbf{steps} and \textit{phases} whose parameters can all be set or modified before execution.
We now describe, in the next \num{4} sections, the concepts of \textbf{steps}, \textit{phases} and \textbf{systems} around which ArcaNN is organized and address what user files are needed to start the ArcaNN procedure.

\subsubsection{Steps}
\begin{figure}[h]
    \centering
    \includegraphics[width=8.3cm]{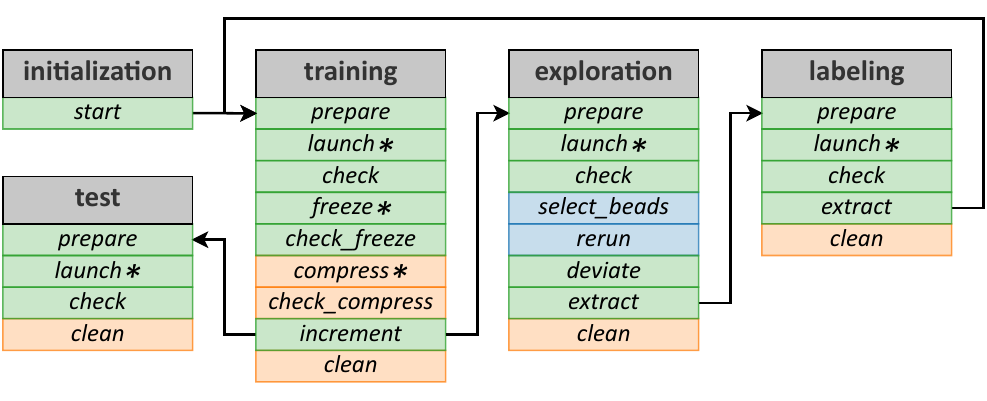}
    \caption{
    The ArcaNN's architecture divided into five main modules corresponding to the successive \textbf{steps}: initialization, training, exploration, labeling and an optional test module.
    Each module is divided into several \textit{phases} that are executed in sequence with user intervention, either proceeding from top to bottom within the same module or by following the arrows between modules.
    \textit{Phases with an asterisk (*) invoke a scheduler to submit the resource-intensive jobs to the HPC, while the others are almost instantaneous and are executed on the login node.}
    In green are the \textit{phases} that are mandatory, in orange the optional \textit{phases} and in blue the \textit{phases} that are mandatory only in the case of path-integral MD simulations.
    }
    \label{fig_main:arcann}
\end{figure}

ArcaNN's architecture is structured around five modules (each corresponding to a \textbf{step} in the concurrent learning scheme, see Figure \ref{fig_main:nnp_concurrent_learning_arcann}B): initialization (1), training (2), exploration (3), labeling (4), and testing (5) (Figure \ref{fig_main:arcann}).
Each \textbf{step} is further divided into a succession of \textit{phases}, which are detailed below.

\subsubsection{Phases}
A \textit{phase} is a subdivision, a specific execution of ArcaNN within a \textbf{step}, and corresponds to the command the user executes.
The outcomes of each \textit{phase} within a \textbf{step} are stored in JSON files in a \textit{control} folder, easily readable by the user.
This ensures the traceability of the workflow, and allows retrieving information in an automated way.
In particular, the status of each \textit{phase} within a \textbf{step} is recorded and checked, avoiding the risk of skipping a non-optional \textit{phase} or doing them in the wrong order.
In addition, from iteration to iteration, if no user input is provided, parameters are propagated or re-calculated automatically.

ArcaNN requires minimal user input beyond the user files detailed below and a comprehensive manual accompanied with example files is provided on the GitHub repository\cite{davidArcaNN2024}.
ArcaNN generates all the necessary files and directories for the workflow; its operating parameters are set to default values unless tuned on demand by the user.
Each time a \textit{phase} is executed, two JSON files are created.
One is the default JSON (\textit{default\_input.json}), where all the default values used are stored, providing guidelines for the user.
The other JSON (\textit{used\_input.json}) stores all the values used for this specific \textit{phase} and is created only if the \textit{phase} is successfully completed, ensuring the traceability of the values used for each \textit{phase} in each iteration.

Any default value can be modified by the user-provided JSON file (\textit{input.json}).
For example, if a user executes the \textit{training prepare} phase but wants to change the learning rate, they can provide an \textit{input.json} containing only the learning rate value.
The user will then relaunch the \textit{training prepare} phase and the \textit{input.json} values will be taken into account.
In this scheme, the priority is given to user values, then to values used in the previous iteration or auto-calculated from the previous iteration, and finally to the default values.
This is useful, for example, if the user wants to change a parameter in the \textbf{exploration}; this change will be carried over to the next iteration without the need to provide an input file again.
If a \textit{phase} fails, an explicit message will be displayed, and the user will have to fix the issue before re-executing the \textit{phase}.

\subsubsection{Systems}

A training dataset for a MLIP should be representative of the chemical phase space, and can include configurations with different chemical compositions, different thermodynamic conditions, and different exploration biases.
In ArcaNN, this is described by \textbf{systems}.
Each \textbf{system} is characterized by its chemical composition (\textit{e.g.}, reactant at different concentrations), thermodynamic conditions (temperature or pressure for example), whether the exploration is done with classical or path-integral MD and, if desired, the type of biased sampling along pre-determined CVs that will be executed.
These \textbf{systems} are defined by the user and will be the core of the exploration phase, capital for the generation of the training dataset.
For example, in the process of building a reactive dataset to describe a given chemical reaction from \textit{A to B}, the user could configure twelve \textbf{systems}: (1) unbiased MD simulations of the reactants; (2) unbiased MD simulations of the products; (3) MD simulations starting from the reactants using On-the-fly Probability Enhanced Sampling\cite{invernizziRethinkingMetadynamicsBias2020} (OPES) along one or several CVs that could be good reaction coordinates (RCs); (4) the same simulations starting from the products; (5) steered MD simulations\cite{grubmullerLigandBindingMolecular1996} along similar coordinates, transforming the reactant state into the product state; and (6) the opposite, from product to reactant.
Then, each of these six setups (\num{1}-\num{6}) could be executed at two different temperatures: \SI{300}{\kelvin} and \SI{325}{\kelvin}, leading to a total of twelve \textbf{systems}.

Another feature of ArcaNN is its flexibility: the practical number of \textbf{systems} a user can define depends on their available HPC resources, rather than being constrained by the ArcaNN methodology itself.
Importantly, the \textbf{systems} do not need to have the same chemical composition.
For instance, one might include a \textbf{system} composed of one set of reactants and another that simulates a higher concentration with two sets of reactants, possibly within a larger solvent box.
Furthermore, \textbf{systems} can be constructed from different molecular configurations, such as one with reactants and products, and another with reactants and different products, representing competitive reactions, or even varying solvents to explore a wide range of chemical environments.

\subsubsection{Required user files}

To start the ArcaNN procedure, users should provide two sets of files.
A first set of files corresponds to the user files, which are organized in a \textit{user\_files} folder, with a minimal folder structure as shown in Figure \ref{fig_si:tree_folder}.
We provide skeleton user files in the GitHub repository\cite{davidArcaNN2024} that users can use as a template to create their own files and refer interested users to the documentation for details about these files, including which parts of each file are important to retain so that ArcaNN can read them and auto-fill the needed values.
This choice was made to ensure that users have full control over the files and can adapt them to their needs, as well as to ensure that the framework remains as flexible as possible.
These include the various inputs of the external software used in the workflow, such as CP2K, DeepMD-kit, LAMMPS, i-PI, and Plumed, together with the job scheduler files needed to submit the external software jobs.
It is important to note that, except for the training \textbf{step}, users should provide one input file for each \textbf{system} they wish to simulate, \textit{i.e.} one LAMMPS (or i-PI) input file, one Plumed input file (if needed), one LAMMPS datafile, and one CP2K input file per \textbf{system}.
One LAMMPS datafile is needed per \textbf{system} to define the initial configurations to be simulated.
LAMMPS datafiles are preferred since their format is standardized and more flexible than XYZ files.
A properties file is also needed to specify atomic types and masses.

To control the use of HPC resources, ArcaNN uses a \textit{machine.json} file where HPC resources are identified by a keyword, and the configuration outlines various attributes of the HPC machine, such as the job scheduler, the maximum number of jobs in the queue, and the maximum scheduler array size.
Furthermore, it provides specifics for project or task setups within this HPC resource under sub-keyword, including names for projects and allocations, architecture type, and a designated partition for job queuing as well as valid tasks for execution.
Importantly, it facilitates the incorporation of multiple HPC machines, for executing specific tasks on GPUs and others on CPUs.
An example of a \textit{machine.json} file can be found in Figure \ref{fig_si:JSON_machine}, and more details can be found in the documentation on the GitHub repository\cite{davidArcaNN2024}\textcolor{purple}.

A second set of files required to initiate the training process corresponds to the initial training dataset.
In the current version of ArcaNN, these datasets, which include atomic configurations, energies, forces, and virial tensors, should be formatted for DeePMD-kit.

We pause to provide some useful guidelines on how to generate these datasets.
They are typically obtained from short aiMD simulations.
To enhance the training efficacy, it is recommended that these datasets contain as many uncorrelated configurations as possible, primarily spaced over time.
As a rule of thumb, configurations spaced by \SI{20}{\femto\second} provide a good starting point.

If the aiMD simulations are performed at the same DFT level as the desired reference for the NNPs, the user can directly supply the associated energies, forces, and virial tensors.
However, to improve the computational efficiency, a recommended practice is to conduct the aiMD at a less computationally demanding level of theory before executing the reference level calculations solely on the selected configurations.
This approach is advantageous, as the geometries generated by a cheaper theory level are usually reliable, but the energies and forces are not as good as those provided by a higher level description.
For instance, initial simulations can employ a GGA functional with a minimal basis set, while subsequent reference calculations use a higher level GGA or hybrid functional accompanied by a larger basis set.
Alternatively, users may opt for even more cost-effective calculations, such as semi-empirical methods like DFTB2\cite{elstnerSelfconsistentchargeDensityfunctionalTightbinding1998,elstnerDensityFunctionalTight2014} or GFN2-xTB\cite{bannwarthGFN2xTBAnAccurateBroadly2019}, and then perform the reference calculations on the selected configurations.
ArcaNN offers flexibility in managing the initial datasets, including the option to discard them if their energy distribution significantly deviates from that of the datasets constructed during the iterative training process.
Moreover, it accommodates the addition of extra datasets, independent of the initial and iterative ones, at any stage of the training.
This feature is particularly useful if users provide datasets from other sources or systems that they wish to incorporate.
For example, as initial datasets, users can provide datasets sampling the reactants, the products, and the pathways from reactants to products, and from products to reactants datasets obtained from aiMD.

\subsection{Workflow}

As shown in Figure \ref{fig_main:arcann}, the workflow is divided into five main \textbf{steps}: initialization, executed once at the beginning of the workflow; training (of the NNPs); exploration (swarms of enhanced sampling trajectories with selections of candidates); labeling (labeling the new candidates with the reference method), which are integral parts of the concurrent learning cycle; and testing, which is optional and can be used to assess the training of the NNPs.
A recurrent \textit{phase} is the optional \textit{clean} phase that can be executed at the end of each \textbf{step} to remove unnecessary files, such as temporary files created by ArcaNN and redundant files.
The other phases are specific to each \textbf{step}, and are detailed below.
The next sections will describe the different \textbf{steps} and \textit{phases} of the workflow.

It is important to note that the execution of these steps is not automated; each phase must be manually initiated by the user.
While resource-intensive tasks, such as training, exploration, and labeling, are submitted to the HPC queue manager (e.g., SLURM) for execution, ArcaNN does not provide automatic updates on their completion.
Instead, the user should manually check the status of these tasks in the corresponding check phases before moving on to the next phase.
This method requires more user involvement but ensures precise control over the workflow and facilitates troubleshooting and adjustments based on intermediate results.

\subsubsection{Initialization}

The first \textbf{step} of the workflow is the initialization \textbf{step}, which is executed only once at the beginning of the workflow.
It consists in one \textit{user set-up} phase and an \textit{initialization start} phase.
To initiate the process, users are required to supply a set of initial files to ArcaNN (see above), which are used to generate all the files and directories needed for the subsequent training, exploration, and labeling \textbf{steps}.
After this initial set-up is completed, no additional user-provided files are needed.

When the set-up is complete, the user can proceed to the \textbf{initialization} step which involves a single phase, \textit{start}, ensuring the presence of all the user files.
This step corresponds to the creation of the initial training folder and the \textit{control} directory, where the JSON files are saved.
Additionally, it locates the initial datasets and tags them for the first training step.
This phase also reads all the names of the LAMMPS datafiles provided by the user and then automatically creates the list of \textbf{systems} that ArcaNN will use for the exploration and labeling steps.
In this step, the user can also choose the number of NNP models to train for the committee, which is set to three by default.
After this step is successfully completed, the user can proceed to the \textbf{training} step.

\subsubsection{Training}

This section describes the \textbf{training} step.
The goal of this step is to train a generation of NNPs on the current dataset, and to prepare them for the \textbf{exploration} step.

During the \textbf{training} step, a committee of several NNPs are trained based on the existing structures in the current dataset.
This step is divided into the following phases: \textit{prepare}, \textit{launch}, \textit{check}, \textit{freeze}, \textit{check\_freeze}, \textit{compress}, \textit{check\_compress}, \textit{increment} and \textit{clean}, with an overview of the phases represented in Table \ref{tab:tab_train}.

\begin{table}[!ht]
    \centering
    \caption{Table summarizing the phases of the \textbf{training} step}
    \label{tab:tab_train}
    \begin{tabular}{|p{3cm}|p{10cm}|p{2cm}|}
        \hline
        \textbf{Phase} & \textbf{Description} & \textbf{Status}\\
        \hline
        \textit{prepare} & Create necessary folders and files for the training of the NNPs (and the number of NNPs to be trained)& \textcolor{tabgreen}{Mandatory} \\
        \textit{launch} & Submit training jobs & \textcolor{tabgreen}{Mandatory} \\
        \textit{check} & Check if the training jobs are successful & \textcolor{tabgreen}{Mandatory} \\
        \textit{freeze} & Freeze the NNPs & \textcolor{tabgreen}{Mandatory} \\
        \textit{check\_freeze} & Check if the freezing is successful & \textcolor{tabgreen}{Mandatory} \\
        \textit{compress} & Compress the NNPs & \textcolor{taborange}{Optional} \\
        \textit{check\_compress} & Check if the compression is successful & \textcolor{taborange}{Optional} \\
        \textit{increment} & Update the temporary number & \textcolor{tabgreen}{Mandatory} \\
        \textit{clean} & Remove unnecessary files & \textcolor{taborange}{Optional} \\
        \hline
    \end{tabular}
\end{table}

The \textit{prepare} phase will create the necessary folders and files for the next phase.
It will copy the datasets, and the dptrain.json (which is the DeePMD-kit input) file to the training folder and we refer to the documentation of DeepMD-kit\cite{zengDeePMDkitV2Software2023} for this file and the associated hyperparameters.
In this phase the user can define, for example, the learning rate, the number of steps, and the machine keyword for the job scheduler parameters (for more details, see the documentation on the Github repository\cite{davidArcaNN2024}).

All the subsequent phases do not require further user inputs.
After the \textit{prepare} phase, the \textit{launch} phase will submit the training jobs to the HPC cluster.
The \textit{check} phase will check if the training is successful, and will provide guidelines about the training duration that can be used for the next iteration.
The next phase, the \textit{freeze} phase, will submit jobs to the HPC cluster to convert (\textit{i.e.}, freeze) the models from their trainable parameters (\textit{e.g.}, weights and biases) to constants and remove unnecessary training operations, enabling them to be efficiently used for inference (\textit{i.e.}, as NNPs predicting energies and forces), while the \textit{check\_freeze} phase will check the success of this operation.
The \textit{compress} phase will submit jobs to the HPC cluster to compress the models, and the \textit{check\_compress} phase will check the success of compression.
The model compression\cite{luDPCompressModel2022} is used to boost the efficiency of inference using three techniques: tabulated inference, operator merging, and precise neighbor indexing.
This is optional, and the user can choose to skip this phase.
The final phase is the \textit{increment} phase, which updates the iteration number, concluding the active learning cycle by having produced a new generation of NNPs (or the first one).
Figure \ref{fig_si:JSON_training} shows a typical JSON output from this \textbf{step}, located in the \textit{control} folder and named \textit{training\_ITERATIONNUMBER.json}, which records the results of each \textit{phase}.
After the \textbf{training} step is successfully completed, the user can proceed to the \textbf{exploration} step.

\subsubsection{Exploration}

This section details the \textbf{exploration} step and its goal: exploring the chemical space and selecting new candidates.
This is done by propagating (biased) MD simulations with the current NNP generation, then performing a query-by-committee to select and extract inadequately described configurations.

The \textbf{exploration} step is at the core of the construction of a dataset using active learning in order to include representative structures potentially present along the reaction pathway(s).
If the nuclei are treated classically, the current implementation calls LAMMPS for the exploration step, which is divided into the following phases: \textit{prepare}, \textit{launch}, \textit{check}, \textit{deviate}, \textit{extract}, and \textit{clean}.
In the case of quantum nuclei, the exploration is performed using i-PI and is divided into the following phases: \textit{prepare}, \textit{launch}, \textit{check}, \textit{select\_beads}, \textit{rerun}, \textit{deviate}, \textit{extract}, and \textit{clean}.
The overview of the phases is represented in Table \ref{tab:tab_explo}.

\begin{table}[!ht]
    \centering
    \caption{Table summarizing the phases of the \textbf{exploration} step, with the additional mandatory phases for PIMD exploration in blue}
    \label{tab:tab_explo}
    \begin{tabular}{|p{3cm}|p{10cm}|p{2cm}|}
        \hline
        \textbf{Phase} & \textbf{Description} & \textbf{Status} \\
        \hline
        \textit{prepare} & Create necessary folders and files for the exploration (per \textbf{system})& \textcolor{tabgreen}{Mandatory} \\
        \textit{launch} & Submit exploration jobs & \textcolor{tabgreen}{Mandatory} \\
        \textit{check} & Check if the explorations are successful & \textcolor{tabgreen}{Mandatory} \\
        \textit{select\_beads} & Select one random bead per configuration & \textcolor{tabblue}{Mandatory} \\
        \textit{rerun} & Calculate the model deviation on those beads & \textcolor{tabblue}{Mandatory} \\
        \textit{deviate} & Select new candidate configurations & \textcolor{tabgreen}{Mandatory} \\
        \textit{extract} & Extract those configurations & \textcolor{tabgreen}{Mandatory} \\
        \textit{clean} & Remove unnecessary files & \textcolor{taborange}{Optional} \\
        \hline
    \end{tabular}
\end{table}

The \textit{prepare} phase creates the necessary folders and files to run MD simulations for each system using the concurrent NNPs trained at the previous step.
The user can tune the number of trajectories to be run for each NNP (default value of 2).
For example, for six \textbf{systems}, three NNPs, and two trajectories per NNP, a total of $n_{\text{systems}} \times n_{\text{NNPs}} \times n_{\text{trajectories}} = 6 \times 3 \times 2 = 36$ MD simulations will be prepared.
Other tunable parameters include the timestep, the number of steps, and the machine keyword for the job scheduler parameters (see complete list in the repository\cite{davidArcaNN2024}).

The \textit{launch} phase will submit the MD simulations to the HPC cluster, and the \textit{check} phase will ensure the success of the simulations.
If some simulations have crashed, the user can choose to skip them, or to force the selection of candidates along the stable part of the trajectory.
Indeed, it is very common in the early iterations that simulations will crash before the end when encountering structures far from those on which they were trained.
However, they can still be used to enrich the training database.
During this phase, while the MD engine will propagate the trajectory using one of the NNPs, forces are also calculated on-the-fly with the other NNPs.
For a given configuration $x$, the maximal deviation on the atomic forces, $\text{max}_i[\epsilon_{\mathbf{F_i}}(x)]$ is calculated as the maximal deviation on any atom $i$ within the configuration.
The deviation of the atomic forces on atom $i$ for configuration $x$, calculated over the $N$ NNPs, is defined as:

\begin{equation}
    \epsilon_{\mathbf{F},i}(x) = \sqrt{\frac{1}{N} \sum_{k=1}^N || \mathbf{F_{i}}(x,\mathrm{NNP}_{k}) - \langle \mathbf{F_{i}}(x,\mathrm{NNP}_{l}) \rangle_{l=1...N} || ^{2}}
\end{equation}

During the \textit{deviate} phase, configurations are classified into three categories.
Set A includes configurations that closely resemble parts of the training dataset and show minimal variance in the forces, $\text{max}_i[\epsilon_{\mathbf{F},i}(x)] \leq \sigma_{low}$.
Set B includes configurations that present a significant variance in forces, $\sigma_{low} < \text{max}_i[\epsilon_{\mathbf{F},i}(x)] \leq \sigma_{high}$.
Finally, set C includes configurations that are considered as potentially non-physical and unreliable with $\text{max}_i[\epsilon_{\mathbf{F},i}(x)] > \sigma_{high}$.
Configurations within set B will be referred to as candidates and will be labeled and added to the training dataset whereas configurations in set C will be discarded.
The user can modify the values of $\sigma_{low}$ and $\sigma_{high}$, defining the range of set B.

We pause to discuss some useful guidelines for these values gained by our own experience.
We recommend using a $\sigma_{low}$ of about four times the value of the NNP RMSE, which is typically around \SI{0.05}{\electronvolt\per\angstrom}.
Therefore, a value of \SI{0.2}{\electronvolt\per\angstrom} is a good starting point.
Next, $\sigma_{high}$ can be set to four times this value, \textit{i.e.},  \SI{0.8}{\electronvolt\per\angstrom}.
At the later stage of the iterations, the user can reduce these values to \SI{0.1}{\electronvolt\per\angstrom} and \SI{0.4}{\electronvolt\per\angstrom} in order to limit the number of selected configurations once the dataset becomes rich enough in reactive structures.
A third value, $\sigma_{max}$, acts as a threshold beyond which, even if configurations encountered afterwards during the dynamics drop below $\sigma_{high}$, they will still be discarded as the path to these configurations is deemed unphysical, with a default value of \SI{1.0}{\electronvolt\per\angstrom}.
The user can also set the maximum number of candidates to select, which is set to \num{50} by default (for each \textbf{system}), and also set how many timesteps are ignored at the beginning of each trajectory to ensure proper decorrelation from the starting point.

The \textit{deviate} phase also selects starting points for the exploration step of the next iteration.
These are chosen to be the configurations with the lowest deviation in set B, or, in the absence of such candidates, as the last configuration of the dynamics (which belongs to set A).
If no new candidate emerges due to simulations crashing, the starting points of the explorations of the next iteration will be the same as in the current iteration.
This ensures that the next starting points are either part of the training dataset (because they will be candidates belonging to set B) or already well described by the NNPs (set A).
Users also have the option to always start from the same initial configurations, which can be useful at the beginning of the iterative cycle.

The \textit{extract} phase then extracts from the trajectories the selected starting points and candidates for the next step by reading the list of indices from the \textit{deviate} phase.
As the retrieval process can be time-consuming (on the order of minutes, especially if the trajectory files are large), the selection of candidates is split into two \textit{phases}: the first (\textit{vide supra}) is fast as only the deviation files are read, and the user can fine-tune the parameters (the $\sigma$ or the maximum number of candidates) and only then proceeds to the \textit{extract} phase to process the trajectory files and retrieve the candidates configurations.
Furthermore, users have the option to increase the number of candidates twofold by slightly shifting the positions of the atoms\cite{youngTransferableActivelearningStrategy2021}, either for a specific set of atoms or for all atoms.
This process applies to all original candidates, resulting in a final number of candidates that includes both the original and the altered ones.
This is done using the built-in function of Atomsk\cite{hirelAtomskToolManipulating2015} to disturb atomic positions by applying random translation vectors to atoms, while ensuring no global translation of the system and following a normal distribution function to generate new configurations.
This can be useful when the exploration phase does not yield enough candidates or if the user wishes to explore a wider range of the phase space.
Caution is emphasized, as the disturbed move is done randomly and could lead to unphysical configurations, and can be also time-consuming if the number of candidates is large.

ArcaNN also offers the possibility to train the NNPs for nuclear quantum effects using RPMD and in that case, path-integral MD are run with i-PI.
To have accurate NNPs to perform the RPMD simulations, they are trained on the beads and not on the centroids, as the NNPs will be used to compute forces on each bead.
It is possible to use NNPs trained on PIMD simulations to perform classical MD simulations as the classical nuclei lie between beads thus the NNPs can interpolate the computed forces (and energies), but the beads cannot be reliably extrapolated from a training on classical nuclei, thus caution is advised in the latter case.
To achieve this, the \textbf{exploration} step has two new phases, \textit{select\_beads} and \textit{rerun}.
As i-PI does not allow multiple models to calculate the model deviation on-the-fly, the \textit{select\_beads} phase will randomly select one bead per MD step, and the \textit{rerun} phase will run inference on the 'trajectory' to get the deviation between the models using LAMMPS.
The user can also mix classical and path-integral MD simulations, with one set of \textbf{systems} for each type of simulation.

It is important to note that most of the exploration (during the \textit{prepare} phase) and selection (during the \textit{deviate} phase) parameters, as well as the possibility to create new perturbed configurations (during the \textit{extract} phase), are set independently for each \textbf{system}, providing great flexibility to the user, who can either use the same values for all systems or set different values for each system.
Identically to the \textbf{training} step, a JSON output is written, and an example is shown in Figure \ref{fig_si:JSON_exploration}.

A key point is that if the previous iteration $N$ results in a limited pool of candidates, ArcaNN dynamically adjusts the MD simulations lengths for the following exploration phase $N+1$, aiming to increase the sampling.
After the \textbf{exploration} step is successfully completed, the user can proceed to the \textbf{labeling} step.

\subsubsection{Labeling}

This section describes the \textbf{labeling} step.
It will present the methods used to label the new candidates selected in the exploration step, which will then enrich the training dataset.

The goal of this \textbf{step} is to generate labels for the candidates selected in the exploration step, which will then enrich the training dataset.
This step is divided into several phases: \textit{prepare}, \textit{launch}, \textit{check}, \textit{extract}, \textit{clean} and a overview of the phases is represented in Table \ref{tab:tab_labeling}.

\begin{table}
    \centering
    \caption{Table summarizing the phases of the \textbf{labeling} step}
    \label{tab:tab_labeling}
    \begin{tabular}{|p{3cm}|p{10cm}|p{2cm}|}
        \hline
        \textbf{Phase} & \textbf{Description} & \textbf{Status} \\
        \hline
        \textit{prepare} & Create necessary folders and files for the labeling of the candidates & \textcolor{tabgreen}{Mandatory} \\
        \textit{launch} & Submit the labeling jobs & \textcolor{tabgreen}{Mandatory} \\
        \textit{check} & Check if the labeling jobs are successful & \textcolor{tabgreen}{Mandatory} \\
        \textit{extract} & Extract the labeled candidates & \textcolor{tabgreen}{Mandatory} \\
        \textit{clean} & Remove unnecessary files & \textcolor{taborange}{Optional} \\
        \hline
    \end{tabular}
\end{table}

As with the other steps, the \textit{prepare} phase will ensure the creation of necessary folders and files to run the single-point calculations.
A few options are available to the user besides providing the input files for CP2K, namely the number of nodes, the number of MPI processes per node, as well as the number of threads per MPI process.
To improve efficiency, the single-point (SP) calculations are divided into two parts: the first SP calculation can be a quick and cheap calculation (\textit{e.g.}, GGA with a small basis set) to get an initial optimized wavefunction which will serve as a guess for the second SP calculation at the desired reference level of theory (\textit{e.g.}, GGA or hybrid-GGA with a large basis set).
This significantly speeds up the labeling calculation.

The \textit{launch} phase will submit the single-point calculations to the HPC cluster, and the \textit{check} phase will ensure the success of the calculations (\textit{i.e.}, the convergence of the calculations).
If a cheap calculation did not converge, a warning will be displayed; however, if the subsequent expensive calculation did converge, the program will continue.
If the expensive calculation did not converge, an error will be displayed, and the user will have to fix the issue before relaunching the phase, either by skipping the candidate or by manually relaunching the single-point calculation.

The \textit{extract} phase will extract the molecular structure, energy, forces, box size, and, if present, the virial tensor from the single-point calculations and store them in the DeepMD-kit format as a new dataset.
By convention, the files containing these new labeled structures are named \textit{sysname\_XXX}, where sysname is the name of the system and XXX is the iteration number.

As per the previous \textbf{steps}, a JSON file is written and is shown in Figure \ref{fig_si:JSON_labeling}.
After the \textbf{labeling} step is successfully completed, the user can proceed to the \textbf{training} step completing the cycle.

\subsubsection{Test}

An optional step, the \textbf{test} step is used to test the NNPs performances against the reference methodology after each \textbf{training}.
This step is divided into several phases: \textit{prepare}, \textit{launch}, \textit{check}, \textit{clean}, with an overview of the phases represented in Table \ref{tab:tab_test}.

\begin{table}
    \centering
    \caption{Table summarizing the phases of the \textbf{test} step}
    \label{tab:tab_test}
    \begin{tabular}{|p{3cm}|p{10cm}|p{2cm}|}
        \hline
        \textbf{Phase} & \textbf{Description} & \textbf{Status} \\
        \hline
        \textit{prepare} & Create necessary folders and files for the testing of the NNPs & \textcolor{tabgreen}{Mandatory} \\
        \textit{launch} & Submit the testing jobs & \textcolor{tabgreen}{Mandatory} \\
        \textit{check} & Check if the testing jobs are successful and concatenate the results in a JSON file& \textcolor{tabgreen}{Mandatory} \\
        \textit{clean} & Remove unnecessary files & \textcolor{tabgreen}{Mandatory} \\
        \hline
    \end{tabular}
\end{table}

The \textit{prepare} phase will ensure the creation of the necessary folders and files to run the testing phase.
It is important to note that here, the testing is done on all datasets, including the initial, iterative, and extra datasets.
This is not a validation of the NNPs, but a way to ensure that the NNPs are still performing well on all datasets.
For a more in-depth validation, the user should provide a separate dataset they have not used for training.
The \textit{launch} phase will submit the testing jobs to the HPC cluster and the \textit{check} phase will ensure the success of the testing jobs as well as writing the results in a control JSON file (Figure \ref{fig_si:JSON_testing}).

\section{Application to typical chemical reactions}

In this section, we demonstrate the use and capabilities of ArcaNN in training NNPs on two examples: a nucleophilic substitution reaction in solution and a pericyclic reaction in the gas phase. These two reactions are selected as model test cases for which all the necessary files are provided; however, we stress that a prototype version of ArcaNN has been successfully used for more complex sequential reactions involving nucleophilic attack, nucleofuge departure and proton rearrangements\cite{davidCompetingReactionMechanisms2024,benayadPrebioticChemicalReactivity2024}.

\subsection{Nucleophilic substitution reaction}

\begin{figure}[h]
    \centering
    \includegraphics[width=8.3cm]{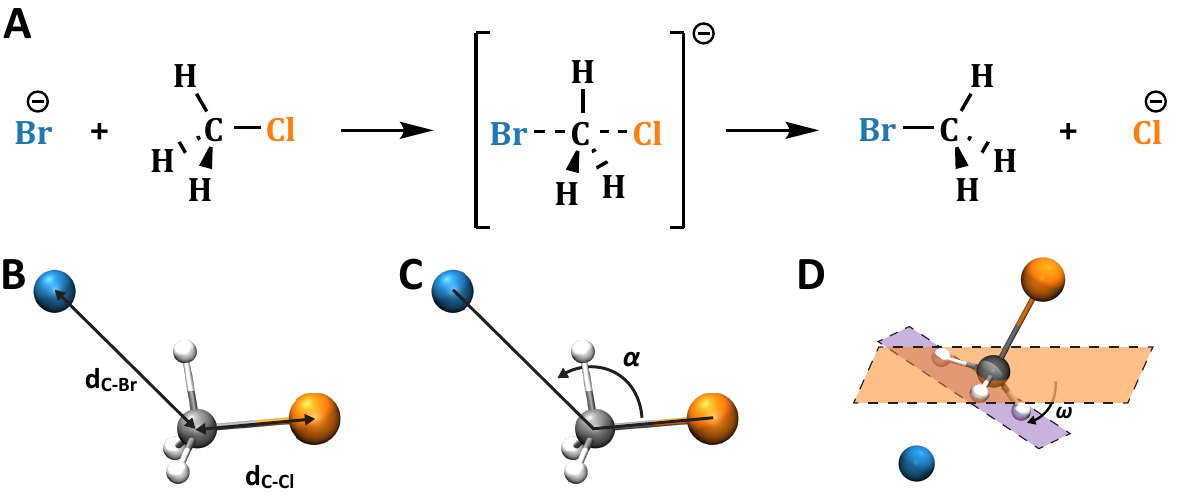}
    \caption{
        (A) Mechanism of the $S_{N}2$ reaction between chloromethane and bromide ion.
        Collective variables used to bias or to monitor the reaction:
        (B) The distances between the carbon atom of the methyl group and the chlorine and bromine atoms, $d_{C-Cl}$ and $d_{C-Br}$, respectively;
        (C) The angle $\alpha$ between the carbon atom of the methyl group and the chlorine and bromine atoms;
        (D) The angle $\omega$ between the plane containing the three hydrogen atoms of the methyl group (purple) and the plane containing the carbon atom of the methyl group and two hydrogen atoms (orange).
    }
    \label{fig_main:reaction_and_cvs}
\end{figure}

We focus on the $S_{N}2$ reaction between chloromethane \ch{CH3Cl} and a bromide ion \ch{Br-} in acetonitrile \ch{CH3CN}, represented in Figure \ref{fig_main:reaction_and_cvs}.
This reaction together with other related $S_{N}2$ reactions have already been studied using a range of methods including mixed QM/MM simulations and \textit{ab initio} molecular dynamics
\cite{chandrasekharSN2ReactionProfiles1984,chandrasekharTheoreticalExaminationSN21985,bergsmaMolecularDynamicsModel1987,hwangSimulationFreeEnergy1988,raugeiInitioMolecularDynamics1999,raugeiMicrosolvationEffectChemical2001,pagliaiCarParrinelloMolecular2003,valverdeFreeEnergyLandscapeN22022}.

The mechanism involves a single step wherein the \ch{Br-} nucleophile attacks the chloromethane electrophilic carbon from the opposite side of the \ch{Cl} leaving group.
The nucleophilic attack and leaving group departure occur concurrently, leading to the inversion of the carbon center stereochemistry.

\subsubsection{Training of the NNPs}

We present here the key steps of our training strategy, and refer the reader to the Methods section and to the SI for technical details.
All the input files, the labeled datasets, and the NNPs at each iteration are provided on the Github so that interested users can reproduce this procedure step by step.

\paragraph{Initial aiMD dataset}

We start from an exploration of the system in the reactant state (\textit{i.e.}, \ch{CH3Cl + Br-}) using a classical force field.
From this trajectory \num{20} snapshots were extracted with half of them having their bromine and chlorine atoms swapped.
Using these as starting points, very short aiMD trajectories were propagated at the DFT BLYP-D3 level.
By extracting structures as decorrelated in time as possible, we generated an initial dataset of \num{1000} configurations, which will be referred to as the $aiMD$ training dataset (see Methods and Supporting Information).

\paragraph{Iterative non-reactive datasets}

We first performed iterations of the exploration, labeling, and training steps (Figure~\ref{fig_main:arcann}).
The goal was to enrich the dataset in structures not well predicted by a given iteration of the NNP, while not explicitly training for reactivity yet.
In practice, we generated a number of systems that allowed scanning the diversity of arrangements between the two molecules in the reactant and product states.
After \num{7} such iterations, we decided to stop this procedure, as the number of new candidates to be included in the dataset became negligible.
We refer to each generation $i$ of datasets (and their corresponding NNPs) as $NRi$ (for non-reactive).
These steps resulted in a modest enrichment of the initial dataset, with a total number of \num{1158} structures in $NR7$.

\paragraph{Exploration of reactive structures}

Finally, we performed \num{5} iterations of the exploration, labeling, and training steps  with now explicit exploration of structures along the reaction pathway.
This was achieved using a variety of systems based on 1D or 2D OPES.
We refer to each generation $i$ of datasets (and their corresponding NNPs) as $Ri$ (for non-reactive).
These steps resulted in a significant enrichment of the initial dataset, with a total number of \num{2313} (\num{1000} + \num{158} + \num{1155} structures) structures in $R5$.
Although some OPES trajectories crashed during the exploration of reactive structures with intermediate datasets, simulations with $R5$ were found to be stable and we thus decided to stop the dataset construction and training after \num{5} steps (Figure \ref{fig_si:r_structures}).

\subsubsection{Validation of the datasets and their corresponding NNPs}

In this section, we will present how the validation of the training datasets was done and show the advantage of using ArcaNN.
We will detail how to assess the quality of the training, which is essential to ensure the reliability of the NNPs in the case of a chemical reaction, using different metrics.

We now discuss the benefits of the ArcaNN approach by comparing a variety of observables along the iterations.
For this purpose, we first constructed a test dataset that is relevant for the chemical reaction by systematically generating \num{1210} structures along the reactive path between the reactant and product basins using Umbrella Sampling (US) simulations with the final $R5$ NNP (see Methods).
Having a test dataset is critical to assess the quality of the training \cite{morrowHowValidateMachinelearned2023,maxsonEnhancingQualityReliability2024}, and it is generally uniformly sampled along all the phase space.
To study a particular reaction, we believe that a test dataset of untrained structures uniformly sampled along the reaction pathway permits ensuring that the accuracy of the NNPs is constant for all relevant reactive structures.
This is even more important if the reaction presents two pathways: both should be described with the same accuracy.
Independently of the ArcaNN procedure, we also performed two types of enhanced sampling "production-like" simulations at each cycle with the resulting NNPs: US and OPES simulations.
We tracked the occurrence of untrustworthy structures in the US simulations and, for both methods, the free-energy surfaces for the reaction.
These are the metrics we used to determine the validity of the NNPs: the RMSE of the forces for an independent test set along the reaction pathway to ensure accuracy and the stability of the NNPs during enhanced sampling.

\begin{figure*}
    \centering
    \includegraphics[width=11.0cm]{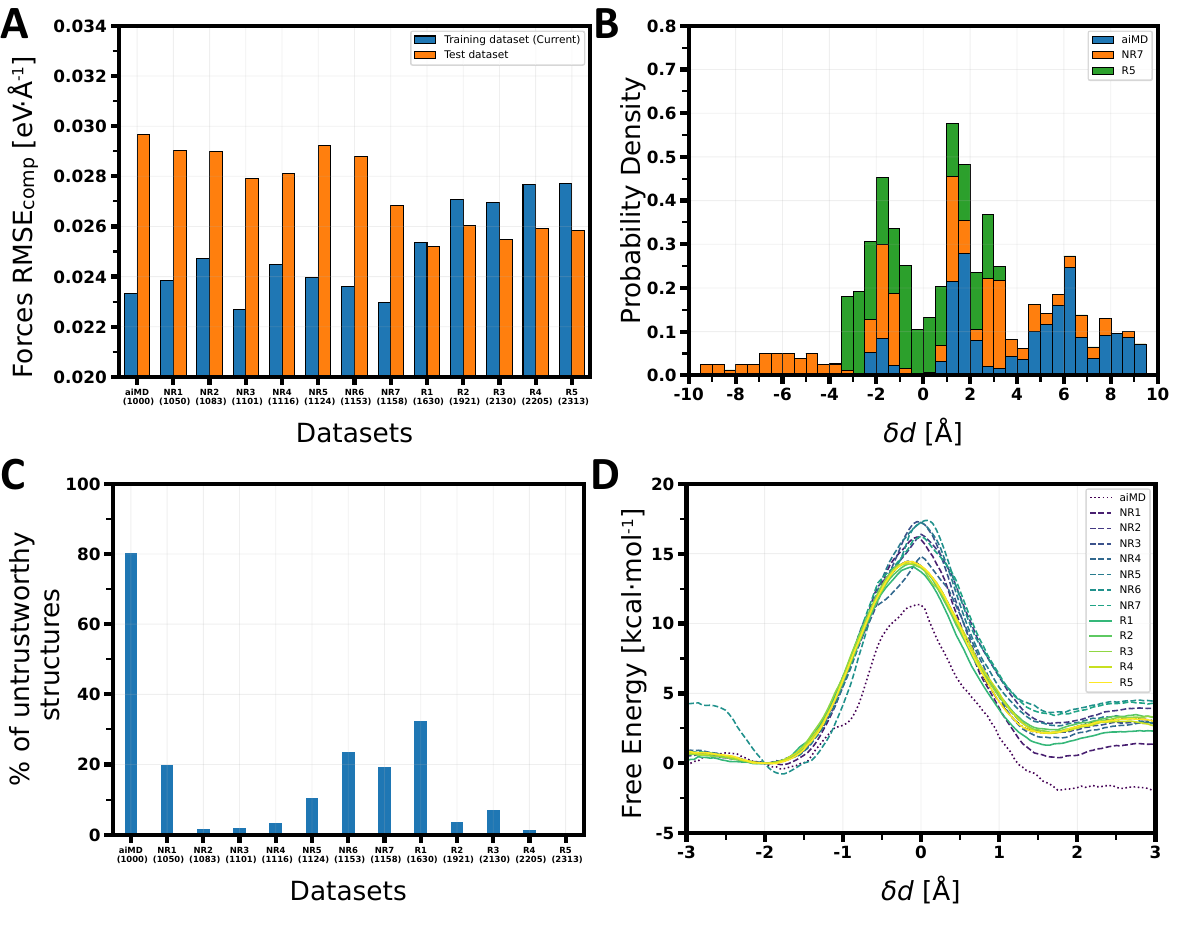}
    \caption{ArcaNN training for the $S_{N}2$ reaction
        (A) Histogram of the RMSE of the forces on the training dataset and the test dataset at each training cycle, with $aiMD$ corresponding to the initial dataset, $NR$ for each non-reactive ArcaNN cycle, and $R$ for each reactive ArcaNN cycle.
        (B) Histogram of the percentage of all untrustworthy structures from the US calculations (where
        $\text{max}_i[\epsilon_{\mathbf{F},i}(x)] >$ \SI{0.7}{\eV\per\angstrom}) with the NNPs obtained at each ArcaNN cycle.
        (C) Probability density of untrustworthy structures in the training datasets as a function of the absolute value of $\delta d$, with $aiMD$ representing the initial structures, $NR7$ representing all \num{158} structures added during the non-reactive ArcaNN cycles, and $R5$ representing all \num{1155} structures added during the reactive ArcaNN cycles.
        (D) Free Energy Profile of the US calculations with the NNPs obtained at each ArcaNN cycle.
    }
    \label{fig_main:errors}
\end{figure*}

Figure \ref{fig_main:errors}A shows the Root Mean Square Error (RMSE) of the force components in the training and test datasets at each ArcaNN cycle.
Until the final iteration of the non-reactive dataset, we do not observe significant variations of the RMSE on the training datasets, suggesting that the NNPs train with similar accuracy, which is not surprising considering the limited augmentation of the training dataset during these iterations.
However, these steps are essential to start mapping the chemical phase space, as the initial aiMD dataset contains a very inhomogeneous distribution of structures, with for example very few reactant configurations where \ch{CH3Cl} and \ch{Br-} are far apart (Figure \ref{fig_main:errors}B).

We notice a clear gap between the RMSE on the training dataset and that on the test dataset that encompasses a lot of reactive structures on which these non-reactive NNPs have not been trained.
However, even without the explicit inclusion of structures on the reaction pathway, the NNPs get better at extrapolating the corresponding forces, leading to a small but noticeable decrease of the RMSE on the test dataset.

When we start reactive cycles, the RMSE on the training dataset suddenly increases, while the error on the testing dataset decreases.
This can be explained by the large number of new structures that are included in the dataset during the reactive cycles, especially close the transition state region (Figure \ref{fig_main:errors}B).
This both degrades the quality of the training but improves the accuracy of the predictions for structures along the reaction pathway, that the NNPs are progressively trained on.

The only observation of the RSMEs can lead to deceptive conclusions about the necessity of iterations and the progressive exploration of the chemical phase space.
Therefore, this should not be the sole aspect to consider to assess the convergence and the quality of the NNPs for a given chemical reaction.
For example, for each generation of NNPs, we report in Figure \ref{fig_main:errors}C the fraction of structures encountered during 1D US simulations (such as those presented in Figure \ref{fig_main:errors}D) that result in large deviations from the reference method.
While the original aiMD NNPs was giving an impression of reasonable RMSEs (Figure \ref{fig_main:errors}A), it results in a dominant fraction of such bad predictions along the reaction pathway.
During the non-reactive cycles, NNPs get progressively better, with $NR2$ and $NR3$ that seem to be reliable.
However, this further degrades again when continuing the non-reactive iterations, which seems surprising since the global RMSE on the test dataset keeps decreasing, although to a limited extent.
This suggests that the non-reactive cycles here could probably have been stopped after the third iteration.

When starting the reactive cycles, the NNPs become more and more reliable when considering the fraction of untrustworthy structures (Figure \ref{fig_main:errors}C and Figure \ref{fig_si:r_structures}), which goes to zero for the fifth iteration $R5$.
However, things do not seem to significantly improve after $R2$.
In Figure \ref{fig_si:rmse_along_pathway}, we represent the RMSEs along the reaction coordinate for the $aiMD$, the $NR7$ and the $R5$ NNPs: one can see that at the final iteration, the RMSEs is constant for all structures encounted along the reaction pathway.
The RMSEs for the $R5$ NNPs and the test dataset are reported in Figure \ref{fig_si:errors_on_forces_training_test}.
The RMSEs on the magnitude of the forces are similar for the training and test datasets with a value around \SI{0.05}{\electronvolt\per\angstrom}, whereas the RMSEs on the forces components are equal and slightly lower, with values around \SI{0.03}{\electronvolt\per\angstrom}.

One key aspect that is overlooked in these considerations is the stability of the NNPs when running the actual simulations, especially so when using enhanced sampling methods.
For example, when running the 1D US simulations for each generation of NNPs, many windows crash after a few tens to a few hundreds of ps.
This is observed for all NNPs except the last one ($R5$).
However, these simulations provide enough data to allow for overlap between adjacent windows along this collective variable, and the corresponding PMFs can be determined (Figure \ref{fig_main:errors}D).
Despite being not stable, the intermediate NNPs lead to free-energy profiles that do not exhibit major inconsistencies, although the barrier appears to be not quantitatively described when the aiMD or non-reactive NNPs are used.
Strikingly, the transitions state (TS) structure is not correct, being a carbocation, as in a $S_N1$ mecanism (Figure \ref{fig_si:fep_cv_us_nr2_nr3}).
For more complex reactions involving several atom exchanges (for example, proton transfers in addition to a heavy atom exchange), it is expected that free-energy surfaces would not easily converge.

\begin{figure}[h]
    \centering
    \includegraphics[width=5cm]{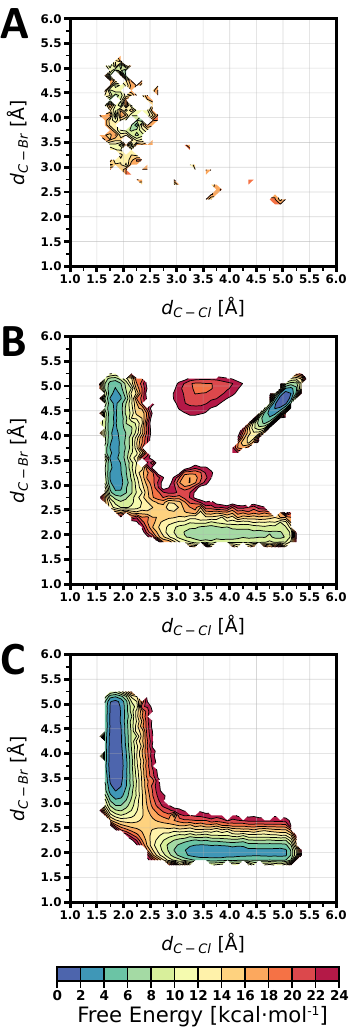}
    \caption{
        Free energy surfaces for the $S_{N}2$ reaction obtained from OPES simulations with the NNPs trained respectively on the $R1$ (A), $R3$ (B) and $R5$ (C) datasets.
        }
    \label{fig_main:opes}
\end{figure}

However, US simulations give seemingly physical results with non reactive NNPs for this specific case, which may fool the user into believing that subsequent optimization of the NNPs are not required.
As already mentioned, the fact that all but the final $R5$ NNPs result in at least one non stable trajectory is already an indication that they should not fully be reliable.
Long enhanced sampling simulations using \textit{e.g.}, OPES appear as a more stringent test of the quality of these NNPs (see Methods for details).

For example, OPES simulations with the NNPs from the $R1$ dataset crash after \SI{44}{\pico\second} and the one on the $R3$ dataset does not crash but starts to be untrustworthy after \SI{978.25}{\pico\second}.
When accounting for the bias accumulated until they crash or become untrustworthy, we can reconstruct free energy surfaces along the carbon-halogen distances (see Figure \ref{fig_main:opes}), which are not correct at all and exhibit unrealistic basins.
Only the final $R5$ NNP converges to a $\Delta G^{\ddagger}$ equal to \SI{14.74 \pm 0.39}{\kcal\per\mol} and a $\Delta G$ equal to \SI{2.25 \pm 0.44}{\kcal\per\mol}, similar to the values obtained from the US simulations with the same NNP (see below).

These results illustrate that the RMSE of the forces on a test dataset is not enough to ensure the validity of the NNPs.
One must also check the stability of the NNPs during enhanced sampling simulations, because the explored pathways are not always the minimum free energy paths and US simulations with very high number of windows and good overlap can mask this instability.
We recommend to use several types of enhanced sampling simulations to ensure the stability of the NNPs, ideally using a superset of those that will be used for the study of the reaction, especially when the reaction require more than one collective variable to be described.

\subsubsection{Thermodynamics and mechanism of the model reaction}

We now present how the final NNPs can be used to study the $S_{N}2$ reaction between chloromethane and bromide ion in acetonitrile.
This will be done using two types of enhanced sampling simulations: Umbrella Sampling (US) and On-the-fly Probability Enhanced Sampling (OPES).

\begin{figure*}
    \centering
    \includegraphics[width=12.7cm]{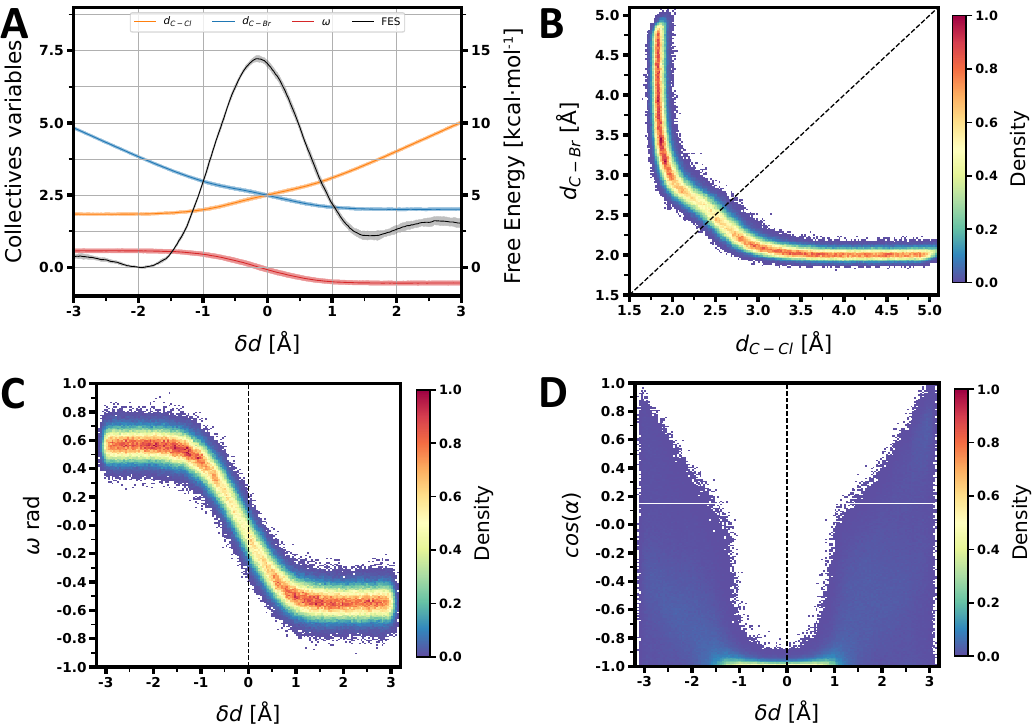}
    \caption{
        (A) Free energy surface of the $S_N2$ reaction obtained from the US simulations with the NNP trained on the $R5$ dataset, with the average value (solid colors) and the 95\% confidence interval (shaded colors) and the average value of the collective variables (as well as the 95\% confidence interval) for each block of the US simulations (shaded colors).
        (B) Joint density distribution of the distance $d_{C-Cl}$ and the distance $d_{C-Br}$, with the dotted line representing $\delta d$ = \SI{0}{\angstrom} obtained from the US simulations.
        (C) Joint density distribution of the $\omega$ angle and $\delta d$ obtained from the US simulations.
        (D) Joint density distribution of the cosine of the $\alpha$ angle and $\delta d$ obtained from the US simulations.
    }
    \label{fig_main:free_energy}
\end{figure*}

We eventually illustrate how the final, stable $R5$ NNP can lead to quantitative and accurate information about this model $S_N2$ reaction.
In Figure \ref{fig_main:free_energy}A, we show the free-energy profile along the asymmetric strech of the carbon-halongen distances $\delta d$, together with the evolution of these distances and of the $\omega$ angle reporting on the Walden inversion.
Figure \ref{fig_main:free_energy}B-D shows some joint probabilities of these key collective variables (CVs) along the reaction.

Based on the free-energy profile, we determined the reaction free energy, directly from the free energy profile, $\Delta G$ to be \SI{2.20 \pm 0.23}{\kcal\per\mol} and the reaction free energy barrier $\Delta G^{\ddagger}$ to be \SI{14.46 \pm 0.17}{\kcal\per\mol}.
The transition state is located at $\delta d$ = \SI{-0.175}{\angstrom}, consistent with an $S_{N}2$ reaction and an associative mechanism as we can see in Figure \ref{fig_main:free_energy}A.
At $\delta d$ = \SI{-0.175}{\angstrom}, the distance $d_{C-Cl}$ and $d_{C-Br}$ are equal to \SI{2.4}{\angstrom} and \SI{2.575}{\angstrom} respectively.

In Figure \ref{fig_main:free_energy}B, the density distribution of the cosine of the $\alpha$ angle, formed by the chlorine atom, the carbon atom, and the bromine atom (see Figure \ref{fig_main:reaction_and_cvs}C), along $\delta d$, is reported.
In both the reactant and product states, $\alpha$ is uniformly distributed at large distances when the two molecules do not interact, but becomes more and more colinear as we approach the transition state, taking a value of \SI{171}{\degree}.
This behavior is expected for the $S_{N}2$ reaction mechanism, where the nucleophile attacks the carbon atom from the opposite side of the leaving group.
The density distribution of the $\omega$ angle defined as the angle between the plane formed by the three hydrogens of the chloromethane and the plane formed by the carbon and two of the three hydrogens of the chloromethane is reported in Figure \ref{fig_main:free_energy}C.
$\omega$ (see Figure \ref{fig_main:reaction_and_cvs}D) takes a value of \SI{32.6}{\degree} in the reactant state and \SI{-30.9}{\degree} in the product state, reaching a value of \SI{2.3}{\degree} at the transition state, demonstrating a Walden inversion\cite{waldenUeberGegenseitigeUmwandlung1896} of chloromethane, characteristic of the $S_{N}2$ reaction.

\subsection{Diels-Alder reaction}

We now illustrate the capabilities of ArcaNN on another type of reaction. We select a pericyclic reaction consisting of a [4 + 2] addition: the Diels-Alder reaction between ethylene (\ch{C2H4}) and 1,3-butadiene (\ch{C4H6}) in the gas phase, forming cyclohexene (\ch{C6H10}) (see Figure \ref{fig_main:da_reaction_cv_error_fep}A).
This reaction has been extensively studied using a wide range of theoretical methods \cite{sakaiTheoreticalAnalysisConcerted2000,domingoUnderstandingMechanismPolar2009,cuiThoroughUnderstandingDiels2014,pestanaDielsAlderReactions2020,youngReactionDynamicsDiels2022}. For simplicity, we focus on the reactivity of the \textit{s}-cis conformation of 1,3-butadiene, which is the most reactive form of the molecule\cite{cuiThoroughUnderstandingDiels2014}

\subsubsection{Training of the NNPs}

In the following, we briefly describe the key steps of the training of the NNPs. Extensive technical details are given in the Methods section and in the SI. Input files and labeled datasets are provided in the Github repository.

\begin{figure*}
    \centering
    \includegraphics[width=12.7cm]{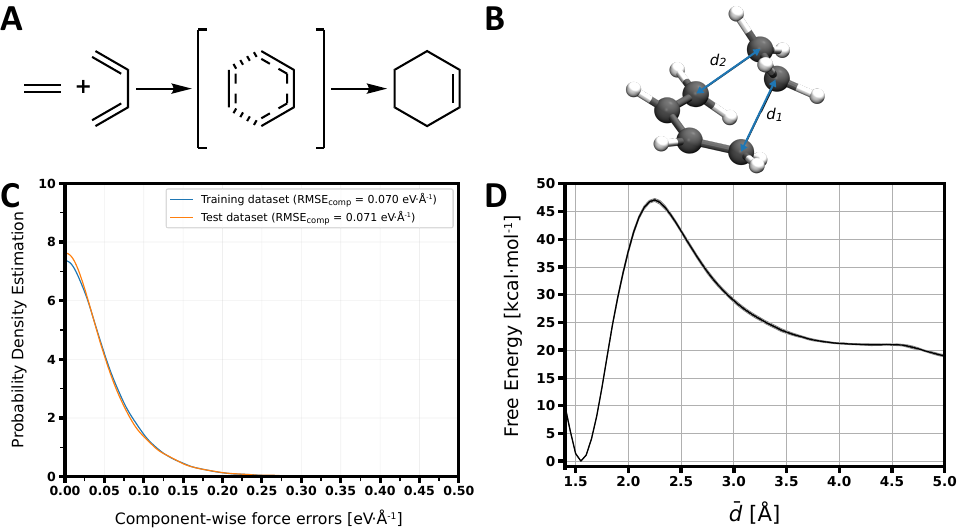}
    \caption{
        (A) Mechanism of the Diels-Alder reaction between ethene and $s$-cis-1,3-butadiene.
        (B) Key collective variables used to describe the reaction: the distances $d_{1}$ and $d_{2}$
        (C) Probability density of the component-wise force errors in the training and test datasets.
        (D) Free energy profile along the average $\bar{d}$ of the $d_1$ and $d_2$ distances obtained from US simulations with the NNP trained on the $R8$ dataset, with the average value (solid colors) and the 95\% confidence interval (shaded colors).
    }
    \label{fig_main:da_reaction_cv_error_fep}
\end{figure*}

The absence of explicit solvent molecules drastically reduces the number of degrees of freedom and hence the training computational complexity.
As a consequence, we directly initiated our training with short aiMD simulations sampling the transition between reactant and product, propagated at the BLYP-D3 DFT level.
We performed one simulation in the reactant state (\ch{C2H4 + C4H6}), with the two molecules kept at close distance, one in the product state (\ch{C6H10}), and two steered-MD simulations along $\bar{d}$, respectively from the reactant to the product and from the product to the reactant.
From these four trajectories, we generated the $aiMD$ training dataset consisting of \num{244} structures (\num{61} structures per trajectory) (see Methods and Supporting Information).

From this initial dataset, we started the ArcaNN procedure with a mixture of non-reactive and reactive systems based on steered-MD and 1D OPES along the average $\bar{d}$ of the two distances ($d_{1}$ and $d_{2}$), corresponding to the newly formed bonds, see Figure \ref{fig_main:da_reaction_cv_error_fep}B.
We performed \num{8} iterations of the exploration, labeling, and training steps, with a total number of \num{3519} (\num{244} + \num{3275}) structures in the final dataset, named $R8$.
The training was considered as converged at this point as very few new structures were added to the dataset during the last iteration (< \num{1}\% of the total number of structures generated during the last exploration).

To assess the quality of the NNPs, we constructed a test dataset of \num{1095} structures along $\bar{d}$ generated using US simulations with the final $R8$ NNP (which were primarily used to calculate the reaction free-energy landscape, see Methods).
In Figure \ref{fig_main:da_reaction_cv_error_fep}C, we show the distribution of errors on the force components in the training and test datasets at the final ($R8$) cycle.
The RMSE of the component of the forces is $\simeq$\SI{0.07}{\eV\per\angstrom} on the training dataset and $\simeq$\SI{0.07}{\eV\per\angstrom} on the test set (see Figure \ref{fig_si:errors_on_forces_training_test_da}); the RMSE on the forces along the reaction pathway is represented in Figure \ref{fig_si:rmse_along_pathway_da}.

\subsubsection{Thermodynamics and mechanism of the model reaction}

Using the final $R8$ NNP, we performed US simulations along the collective variable $\bar{d}$ to calculate the free energy profile of the Diels-Alder reaction (Figure \ref{fig_main:da_reaction_cv_error_fep}D).
The minimum at $\bar{d} =$ \SI{1.5}{\angstrom} corresponds to the product state (\ch{C6H10}), the relatively flat region beyond $\bar{d} =$ \SI{4.85}{\angstrom} corresponds to the reactant state (\ch{C4H6 + C2H4}), and the maximum at $\bar{d} =$ \SI{2.25}{\angstrom} corresponds to the transition state.
The $\Delta G$ and $\Delta G^{\ddagger}$ were calculated from the free energy profile to be \SI{-19.0 \pm 0.2}{\kcal\per\mol} and \SI{28.1 \pm 0.1}{\kcal\per\mol}, respectively.
This is in fair agreement with the work of Cui and Liu\cite{cuiThoroughUnderstandingDiels2014}, who  reported values for $\Delta G$ of \SI{-14.3}{\kcal\per\mol} and $\Delta G^{\ddagger}$ of \SI{33.2}{\kcal\per\mol} using a static approach at the same level of theory.
As per the $S_{N}2$ reaction, we just report the free energy difference between the different states.
We note that our reactant state is not at infinite distance as in the work of Cui and Liu. If we thus examine the better defined $\Delta G^{\ddagger}$ of the reverse process, {\em i.e.} the ring-opening reaction, we find an excellent agreement between our computed value \SI{47.1 \pm 0.2}{\kcal\per\mol} and the previously-published value of \SI{47.5}{\kcal\per\mol}.

For this prototypical Diels-Alder reaction, our simulations suggest that the mechanism is concerted and  quasi-synchronous, with the two bonds forming at the same time, in agreement with the literature\cite{houkEvidenceConcertedMechanism1986,sakaiTheoreticalAnalysisConcerted2000,singletonIsotopeEffectsDistinction2001}.
This can be seen on the probability density distribution of the $d_{1}$ and $d_{2}$ distances along the reaction coordinate $\bar{d}$ (see Figure \ref{fig_si:distribution_d1_d2_US_da}).

\subsection{Methods}

The methods section outlines the generation of initial datasets from aiMD simulations and the subsequent training of NNPs with ArcaNN.
It details the non-reactive and reactive iterative training cycles, including dataset augmentation and parameter settings.
Finally, it describes the production simulations performed using US (and OPES for the $S_{N}2$) simulations to explore system reactivity and calculate free energy profiles.

\subsubsection{Initial datasets}

The initial datasets were generated through ab initio molecular dynamics (aiMD) simulations for both the $S_{N}2$ and Diels-Alder reactions.
For the $S_{N}2$ reaction, twenty trajectories of \SI{2}{\pico\second} each were performed with a timestep of \SI{0.5}{\femto\second}. Ten trajectories started from the reactant state (\ch{CH3Cl + Br-}) and the other ten from the product state (\ch{CH3Br + Cl-}).
In the Diels-Alder simulations, four trajectories of the same length and timestep were conducted: one initiated from the reactant state (\ch{C4H6 + C2H4}) with the molecules in close proximity, another from the product state (\ch{C6H10}), and two steered-MD simulations transitioning between reactant and product states in both directions.
Structures were extracted every \SI{30}{\femto\second} from each aiMD trajectory after discarding the initial\SI{0.5}{\femto\second} to ensure proper decorrelation.
For the $S_{N}2$ reaction, these structures were combined into two sets -- one for the reactant and one for the product -- each containing \num{500} configurations.
In the case of the Diels-Alder reaction, the structures were grouped into four sets corresponding to each trajectory, totaling \num{244} configurations.
All configurations were labeled at the BLYP-D3 level of theory using the TZV2P-MOLOPT basis set for $S_{N}2$ and the TZV2P basis set for the Diels-Alder reaction, along with GTH pseudopotentials for both; this is referred to as the reference level.
The molecular structures, along with their corresponding box sizes, energies, forces, and virial tensors, were extracted and stored in the DeepMD-kit format.
These datasets were then provided as initial inputs for use with ArcaNN, comprising \num{1000} configurations divided into two datasets for the $S_{N}2$ reaction and \num{244} configurations divided into four datasets for the Diels-Alder reaction, collectively referred to as the $aiMD$ training dataset.

\subsubsection{Initialization}

In the $S_{N}2$ case, this step uses \num{6} \textbf{systems} with \num{3} starting from the reactant state and \num{3} from the product state.
For the reactant state \textbf{systems}, one \textbf{system} was without any restraint, one with a flat-bottom restraint on the distance between the carbon atom of the methyl group and the bromide ion ($d_{C-Br} \leq \SI{3.0}{\angstrom}$, with a force constant $\kappa = \SI{5.0}{\kcal\per\mol\per\angstrom\squared}$), and the last one with a moving harmonic bias (steered-MD) on the $d_{C-Br}$ distance from \SI{2.5}{\angstrom} to \SI{10.0}{\angstrom} with a force constant of \SI{1.0}{\kcal\per\mol\per\angstrom\squared}.
For the product state \textbf{systems}, the same three \textbf{systems} were used, but with the $d_{C-Cl}$ distance.
For the Diels-alder, \num{10} \textbf{systems} were used: two in the reactant and product states without any enhanced sampling; two with steered-MD transitioning from reactant to product and vice versa, acting on both $d_{1}$ and $d_{2}$ distances from \SI{3.5}{\angstrom} (\SI{1.5}{\angstrom}) to \SI{1.5}{\angstrom} (\SI{3.5}{\angstrom}) over \SI{10}{\pico\second} with a force constant of \SI{100}{\kcal\per\mol\per\angstrom\squared}; and six using OPES acting on the collective variable $\bar d$, with initial $\sigma = \SI{0.05}{\angstrom}$, a deposition pace of \num{500} timesteps, and $\Delta E$ values of \SI{20}{\kcal\per\mol}, \SI{50}{\kcal\per\mol}, and \SI{70}{\kcal\per\mol}, starting from both reactant and product states.
In both case, all ArcaNN parameters were kept at their default values; \num{3} NNPs were trained for the committee and \num{2} trajectories per NNP for the exploration step.

\subsubsection{Training}

The training was performed with DeepMD-kit \cite{zengDeePMDkitV2Software2023} version \num{2.1}, with an initial learning rate of \num{0.001} and a final learning rate of \num{1e-06}, a decay rate of \num{0.92}, decay steps of \num{5000}, and a total of \num{400000} steps.
The DeepPot-SE scheme was utilized, setting the cutoff for radial and angular information at \SI{6}{\angstrom} and applying a cosine weight function for atoms located beyond \SI{0.5}{\angstrom}.
The embedding neural network that maps the environment matrix to a local embedding matrix contains \num{3} hidden layers with \num{25}, \num{50}, and \num{100} nodes, respectively.
The following fitting neural network that maps the descriptor to the atomic energy contains \num{3} hidden layers with \num{240} nodes each.
The initial and final energy loss prefactors were set to \num{0.01} and \num{1}, respectively, and the force loss prefactors were set to \num{1000} and \num{1}, respectively.

\subsubsection{$S_{N}2$ non reactive exploration}

The initial exploration was performed using LAMMPS, with a timestep of \SI{0.5}{\femto\second}, a total of \num{20000} steps, and a print interval of \num{200} MD steps (\textit{i.e.}, 1\% of the total length).
The simulations were conducted in the NVT ensemble at \SI{300}{\kelvin} with a CSVR thermostat\cite{bussiCanonicalSamplingVelocity2007} and a time constant of \SI{0.1}{\pico\second}.
The maximum deviation on the atomic forces was set to \num{0.15} for $\sigma_{low}$, \num{0.7} for $\sigma_{high}$, and \SI{1.0}{\eV\per\angstrom} for $\sigma_{max}$ as the candidate selection criteria.
At the seventh iteration, only \num{5} candidates were selected out of the \num{36} MD simulations (three NNPs, two per NNP, and six \textbf{systems}), each lasting \SI{400}{\pico\second}.
Therefore, it was decided to restart the ArcaNN procedure with a biased exploration to include reactive structures.
The total number of configurations in the training dataset at this point was \num{1158}, which will be referred to as the $NR7$ training dataset.

\subsubsection{$S_{N}2$ reactive exploration}

The ArcaNN procedure was restarted with an augmented dataset containing the initial \num{1000} aiMD configurations plus the \num{158} configurations generated by the seven non-reactive cycles.
For this new biased iterative training, twelve \textbf{systems} were created, each with a different starting configuration for the exploration step.
Six \textbf{systems} were used to explore the reactivity using OPES from the reactant state, with three \textbf{systems} where the CV was the $\delta d = d_{C-Br} - d_{C-Cl}$ reaction coordinate and OPES parameters were set to a value $\sigma = \SI{0.05}{\angstrom}$, a deposition pace of \num{2000} timesteps, and $\Delta E$ equal to \SI{5}{\kcal\per\mol}, \SI{10}{\kcal\per\mol}, and \SI{20}{\kcal\per\mol}.
For the other three \textbf{systems}, bias was applied to the $d_{C-Br}$ and $d_{C-Cl}$ distances, with initial values of $\sigma = \SI{0.05}{\angstrom}$ for both, a deposition pace of \num{2000} timesteps, and $\Delta E$ equal to \SI{5}{\kcal\per\mol}, \SI{10}{\kcal\per\mol}, and \SI{20}{\kcal\per\mol}.
The same parameters were used for the \num{6} \textbf{systems} exploring the reactivity using OPES from the product state (with \num{3} OPES 1D and \num{3} OPES 2D).
A total of \num{1155} new configurations from these biased explorations were added to the training dataset.
After \num{7} non-reactive cycles and \num{5} reactive cycles, the number of configurations in the training dataset was \num{2313}, and a final training of the NNPs was performed on this $R5$ dataset.
In figure \ref{fig_main:errors}A, we report the cummulative probability density of structures in the training datasets as a function of the reaction coordinate $\delta d$ for the $aiMD$ dataset (\num{1000} structures), the non-reactive dataset $NR7$ (\num{1000} + \num{158} structures), and the reactive dataset $R5$ (\num{1000} + \num{158} + \num{1155} structures).
It can be seen that the transition region is well sampled with only with the addition of the reactive ArcaNN cycles, and that the non-reactive cycles are not enough to sample the transition region (see also Figure \ref{fig_si:distribution_d1_d2_datasets}).

\subsubsection{Diels-Alder reactive exploration}

Using the \num{10} \textbf{systems} described above, the initial reactive exploration was performed using LAMMPS, with a timestep of \SI{0.25}{\femto\second}, a total of \num{20000} steps, and a print interval of \num{200} MD steps (\textit{i.e.}, 1\% of the total length), at \SI{300}{\kelvin} with a CSVR thermostat\cite{bussiCanonicalSamplingVelocity2007} and a time constant of \SI{0.1}{\per\pico\second}.
After \num{8} iterations of the ArcaNN procedure, the final dataset contained \num{3519} configurations, referred to as the $R8$ training dataset.

\subsubsection{$S_{N}2$ production simulations}

Once the iterative training procedure was finished, the reactivity of the system was explored by performing US simulations with the final NNP (\textit{.i.e.} $R5$).
The reaction coordinate was defined as the difference between the distance $d_{C-Cl}$ and the distance $d_{C-Br}$, $\delta d$ (see Figure \ref{fig_main:reaction_and_cvs}B).
The reaction was divided into \num{121} windows, linearly spaced from $\delta d$ = \SI{-3.0}{\angstrom} to $\delta d$ =\SI{3.0}{\angstrom}.
All simulations thereafter were done in the NVT ensemble at \SI{300}{\kelvin} with a timestep of \SI{0.5}{\femto\second} and a CSVR thermostat\cite{bussiCanonicalSamplingVelocity2007} with a time constant of \SI{0.1}{\per\pico\second}.
For each window, the system was brought to an equilibrium state by performing steered-MD to the target value of $\delta d$, lineary over \SI{50}{\pico\second} with a spring constant of \SI{200}{\kcal\per\mol\per\angstrom\squared}.
Then it was further equilibrated for \SI{50}{\pico\second} at the target value and production runs were done for \SI{600}{\pico\second} for each window.
The total accrued simulation time was \SI{50}{\nano\second} and the simulation speed was roughly \SI{6}{\nano\second\per day} on a single GPU.
A test dataset was also generated by taking \num{10} random structures from the each window of the production US simulations totalling \num{1210} structures along the reaction coordinate $\delta d$ and labeling them at the reference level of theory.

The \SI{600}{\pico\second} long production runs were divided into \num{6} blocks of \SI{100}{\pico\second} each, and the 1D free energy profile was calculated for each block using the Weighted Histogram Analysis Method (WHAM)\cite{kumarWeightedHistogramAnalysis1992} with \num{312} bins along $\delta d$.
Then using each block results, the average and the 95\% confidence interval were calculated by setting the free energy at \SI{0}{\kcal\per\mol} at $\delta d$ = \num{-1.95}.
The $\Delta G$ and $\Delta G^{\ddagger}$ were calculated from the averaged 1D free energy profile as the difference between the free energy of the reactant (\ch{CH3Cl + Br-}) and product states (\ch{CH3Br + Cl-}) and the difference between the free energy of the reactant and the maximum (the transition state) of the free energy profile, respectively.
For the collective variables, each structure for all windows (and the full duration) was binned to a grid of $\delta d$ values (same binning as the WHAM procedure), and the average and 95\% confidence interval were calculated for each bin for the $d_{C-Cl}$ distance, the $d_{C-Br}$ distance, and the $\omega$ angle.

For the OPES simulations with the final $R5$ NNP, bias was applied to the $d_{C-Br}$ and $d_{C-Cl}$ distances, with $\sigma = \SI{0.05}{\angstrom}$ for both, a deposition pace of \num{500} timesteps, and $\Delta E$ equal to \SI{20}{\kcal\per\mol}.
The simulation was propagated for \SI{2.5}{\nano\second} in the NVT ensemble at \SI{300}{\kelvin} with a timestep of \SI{0.5}{\femto\second} and a CSVR thermostat \cite{bussiCanonicalSamplingVelocity2007} with a time constant of \SI{0.1}{\per\pico\second} (same as the production US simulations).
The 2D free energy surface was calculated by reweighting the biased simulations along the $d_{C-Br}$ and $d_{C-Cl}$ distances (Figure \ref{fig_main:opes}).
The simulations were divided into \num{5} blocks of \SI{500}{\pico\second} each, and the free energy was calculated for each block by reweighting along the $\delta d$ collective variables, permitting the calculation of an average and 95\% interval 1D free energy profile (see Figure \ref{fig_si:opes_1d}).
The $\Delta G$ and $\Delta G^{\ddagger}$ were calculated as described above for the US simulations.
The same procedure as the US simulations was used to calculate the average and 95\% confidence interval for the $d_{C-Cl}$ distance, the $d_{C-Br}$ distance, and the $\omega$ angle.

\subsubsection{Diels-Alder production simulations}

After completing the iterative training procedure, the system's reactivity was explored using US simulations with the final NNP, denoted as $R8$.
The reaction coordinate was defined as the average of the two distances $d_{1}$ and $d_{2}$, $\bar{d}$ (see Figure \ref{fig_main:da_reaction_cv_error_fep}B).
The reaction was divided into \num{73} windows, linearly spaced from $\bar d$ = \SI{1.4}{\angstrom} to $\bar d$ = \SI{5.0}{\angstrom}.
All subsequent simulations were performed in the NVT ensemble at \SI{300}{\kelvin} with a timestep of \SI{0.5}{\femto\second}, using a CSVR thermostat\cite{bussiCanonicalSamplingVelocity2007} with a time constant of \SI{0.1}{\per\pico\second}.
To keep the system in the s-cis conformation, a flat-bottom restraint was applied to the dihedral angle with a force constant of \SI{100}{\kcal\per\mol\per\radian\squared} to keep it between \SI[parse-numbers=false]{-\pi/2}{\radian} and \SI[parse-numbers=false]{\pi/2}{\radian} .
For each window, the system was equilibrated by performing steered-MD to the target $\bar d$ value over \SI{50}{\pico\second} with a spring constant of \SI{1000}{\kcal\per\mol\per\angstrom\squared}.
This was followed by an additional \SI{50}{\pico\second} equilibration at the target value, and production runs of \SI{600}{\pico\second} for each window.
The total simulation time accrued was \SI{42.6}{\nano\second}, with a simulation speed of approximately \SI{24}{\nano\second\per day} on a single GPU.
The free energy profile, along with its average and 95\% confidence interval, as well as the $\Delta G$ and $\Delta G^{\ddagger}$ values, were calculated using the same procedure as for the $S_{N}2$ reaction.

\subsubsection{$S_{N}2$ test simulations}

At each ArcaNN cycle, US simulations were performed with a similar protocol than the production of the final NNP.
Using $\delta d$ as the reaction coordinate, \num{121} windows were used, with each starting point being the last geometry of the corresponding window on the production US simulations.
The simulations were done in the NVT ensemble at \SI{300}{\kelvin} with a timestep of \SI{0.5}{\femto\second} and a CSVR thermostat\cite{bussiCanonicalSamplingVelocity2007} with a time constant of \SI{0.1}{\per\pico\second} for \SI{200}{\pico\second}.
The spring constant for each harmonic restraint was set to \SI{200}{\kcal\per\mol\per\angstrom\squared} as per the US simulations.
The free energy profile was then calculated using WHAM\cite{kumarWeightedHistogramAnalysis1992}.
For the OPES simulations done with the $R1$ and $R3$ NNP, exactly the same starting point and parameters were used than the $R5$ production one.

\subsubsection{Diels-Alder test simulations}

A test dataset was generated by selecting \num{15} random structures from each window of the production US simulations, totaling \num{1095} structures along the reaction coordinate $\bar d$, and labeling them at the reference level of theory.

\section{Conclusion}

ArcaNN addresses the challenge of generating training datasets for reactive MLIPs.
By combining concurrent learning with advanced sampling techniques, ArcaNN facilitates the exploration of chemically relevant configurations, including high-energy geometries, and integrates classical and quantum nuclear dynamics into a standardized automated workflow.
The framework is designed to be user-friendly and flexible, allowing researchers to easily set up and run ArcaNN to train neural network potentials (NNPs) for their reactive systems.
We illustrated the power of ArcaNN in the context of two different reactions: first, a nucleophilic substitution ($S_N2$) reaction in explicit solvent, and second, a pericyclic reaction in the gas phase.
In both cases, we demonstrated its capabilities in generating accurate and stable NNPs, both in the reactant and product region, but most importantly along the reaction pathway.
Beyond these simple examples, preliminary versions of the code were used by us for much more complex reactions involving several molecular steps and multiple pathways\cite{davidCompetingReactionMechanisms2024,benayadPrebioticChemicalReactivity2024}.
We also note that the training set obtained for these reactions can be used as a starting point to study similar but more complex reactions presenting different nucleophiles and leaving groups for the $S_{N}2$, or functionalized dienes and dienophiles for the Diels-Alder reaction, respectively.
Although an initial aiMD training set will still be needed to ensure a stable initial representation of the functionalized reagents (especially in the presence of new chemical elements), the reactive configurations in the present dataset will vastly accelerate the iterative procedure needed to refine the description of related systems exhibiting similar reactivity.
We also provide guidelines on how to assess the quality of a NNP for a reactive system, suggesting that many aspects should be considered beyond the canonical RMSE on the energies and forces.
Future developments of ArcaNN will include the incorporation of additional selection techniques, expansion to use other MLIPs, integration with different molecular dynamics engines, and support for various quantum chemistry packages for labeling.

Through continuous improvements, ArcaNN aims to facilitate the broader adoption and application of MLIPs in computational chemistry, enabling new advancements in chemical reactivity and catalysis.

\section{Author contributions}

Rolf David: conceptualization, methodology, software, supervision, validation, formal analysis, investigation, visualization, writing-original draft, writing-review and editing.
Miguel de la Puente: software, validation, resources, writing-review and editing.
Axel Gomez: software, validation, resources, writing-review and editing.
Olaia Anton: validation, resources, writing-original draft, writing-review and editing.
Guillaume Stirnemann: conceptualization, funding acquisition, project administration, supervision, writing-original draft, writing-review and editing.
Damien Laage: conceptualization, funding acquisition, project administration, supervision, writing-original draft, writing-review and editing.

\section{Conflicts of interest}
There are no conflicts to declare.

\section{Data availability}
The code for ArcaNN can be found at \url{https://github.com/arcann-chem/arcann\_training}.
An in-depth documentation is available at \url{https://arcann-chem.github.io/arcann\_training}.
The version of the code employed for this study is version 1.
Necessary user files and initial $aiMD$ datasets to start the training of the NNPs for the $S_{N}2$ and the Diels-Alder reactions with ArcaNN are available in the examples section of the GitHub repository: https://github.com/arcann-chem/arcann\_training.

\begin{acknowledgement}
This work was supported by a postdoctoral fellowship to R.D. from PSL OCAV (Idex ANR-10-IDEX-0001-02PSL) and from the European Research Council (ERC) under the European Union's Eighth Framework Program (H2020/2014-2020)/ERC Grant Agreement No. 757111.
The simulations presented here benefited from an access to the French national high-performance computing resources under  allocations A0130707156 and A0130811005 made by Grand Equipement National de Calcul Intensif.
The authors thank the ENS group members and collaborators who have tested the preliminary versions of the code: Zakarya Benayad, Pierre Vieillard, Oscar Gayraud, Pierre Girard, Anne Milet, Meritxell Malagarriga Pérez, Adrián García-Martínez, Ashley Borkowski, Pauf Neupane, Ward H. Thompson, Mohammadhasan Dinpajooh, and Tomonori Hirano.
\end{acknowledgement}

\begin{suppinfo}

\setcounter{figure}{0}
\setcounter{table}{0}
\renewcommand\thepage{S\arabic{page}}
\renewcommand\thefigure{S\arabic{figure}}
\renewcommand\thetable{S\arabic{table}}
\subsection{Details on the initial \textit{ab initio} MD simulations for the $S_{N}2$ reaction}

The first step consisted on the preparation of the initial datasets by generating reactant structures, by \textit{ab initio} MD simulations.
In order to perform the MD simulations, we have to construct an initial structure of the system.
A \SI{15}{\cubic\angstrom} cubic box was constructed with packmol\cite{martinezPACKMOLPackageBuilding2009}, containing one bromide ion, one chloromethane molecule and 38 acetonitrile molecules.
An energy minimization of the system was then performed using the Amber22 software.
The AMBER's built in General Forcefield (GAFF)\cite{wangDevelopmentTestingGeneral2004} parameters were used for the chloromethane, the acetonitrile and the bromide ion, with the AM1-BCC charge model was used to generate atomic charges
Next, the system was heated to a temperature of \SI{300}{\kelvin} in the NVT ensemble for \SI{20}{\pico\second}, and then equilibrated in the NPT ensemble for \SI{200}{\pico\second} and at a pressure of \SI{1}{\bar} and a temperature of \SI{300}{\kelvin}, both with a timestep of \SI{2}{\femto\second}.
From this equilibration, \num{20} snapshots were extracted, with ten of them having their bromine and chlorine atoms swapped (to generate the product structures).
With these initial structures, \textit{ab initio} MD simulations were performed with the CP2K software\cite{kuhneCP2KElectronicStructure2020} at the BLYP\cite{beckeDensityfunctionalExchangeenergyApproximation1988,leeDevelopmentColleSalvettiCorrelationenergy1988} level of theory and with the D3 dispersion correction\cite{grimmeConsistentAccurateInitio2010}.
The DZVP-MOLOPT-SR\cite{weigendBalancedBasisSets2005,vandevondeleGaussianBasisSets2007} basis set was used in conjunction with the GTH pseudopotentials\cite{goedeckerSeparableDualspaceGaussian1996,hartwigsenRelativisticSeparableDualspace1998,krackPseudopotentialsKrOptimized2005}.
Each run was performed within the NVT ensemble at \SI{300}{\kelvin} with a timestep of \SI{0.5}{\femto\second} for \SI{2}{\pico\second}.
The temperature control was enabled by the use of a CSVR thermostat\cite{bussiCanonicalSamplingVelocity2007} with a time constant of \SI{0.1}{\per\pico\second}.
\clearpage

\subsection{Timings of the ArcaNN training}

\begin{table}
    \centering
    \caption{Summary of Timings for the initial aiMD, the training, exploration, and labeling Phases for the $S_{N}2$ reaction}
    \begin{tabular}{c c c c}
        \hline
        \hline
        \textbf{Phase} & \textbf{Hardware Used} & \textbf{Average Time per Cycle} & \textbf{Total Time} \\
        \hline
        Initial aiMD & AMD EPYC 7H12 & - & 26897.4 core.hours \\
        Training & Nvidia V100 SXM2 & 14.67 gpu.hours & 190.76 gpu.hours \\
        Exploration & Nvidia V100 SXM2 & 59.04 gpu.hours & 767.57 gpu.hours \\
        Labeling & Intel Cascade Lake 6248 & 1836.79 core.hours & 23878.21 core.hours \\
        \hline
        \hline
    \end{tabular}
    \label{tab:timings}
\end{table}
\begin{table}
    \centering
    \caption{Summary of Timings for the initial aiMD, the training, exploration, and labeling Phases for the Diels-Alder reaction}
    \begin{tabular}{c c c c}
        \hline
        \hline
        \textbf{Phase} & \textbf{Hardware Used} & \textbf{Average Time per Cycle} & \textbf{Total Time} \\
        \hline
        Initial aiMD & Cascade Lake 6248 & - & 1928.72 core.hours \\
        Training & Nvidia A100 SXM4 & 3.58 gpu.hours & 32.24 gpu.hours \\
        Exploration & Nvidia V100 SXM2 & 27.81 gpu.hours & 250.33 gpu.hours \\
        Labeling & Intel Cascade Lake 6248 & 297.56 core.hours & 2380.45 core.hours \\
        \hline
        \hline
    \end{tabular}
    \label{tab:timings_da}
\end{table}
\clearpage

\subsection{User-provided tree folder structure}
\begin{figure}[!ht]
    \centering
    \includegraphics{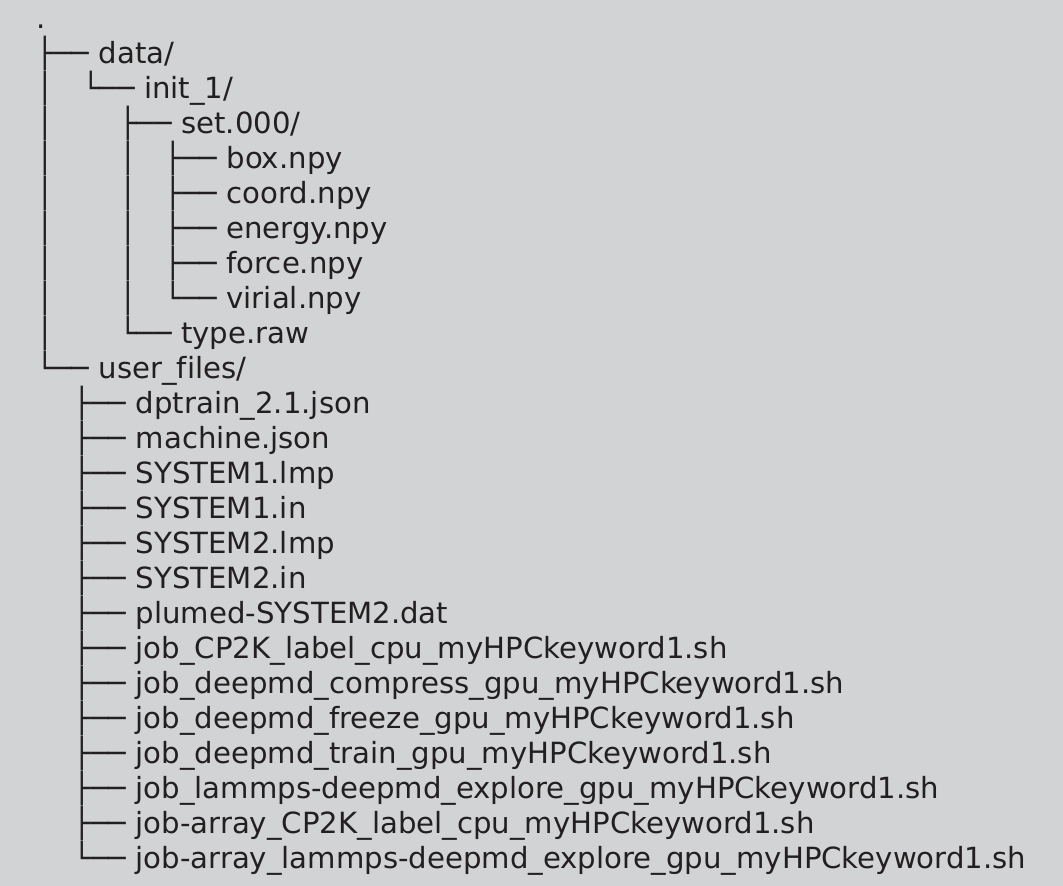}
    \caption{
        Example of the tree folder structure used by ArcaNN.
        }
    \label{fig_si:tree_folder}
\end{figure}
\clearpage

\subsection{Machine JSON user file used by ArcaNN}
\begin{figure}[!ht]
    \centering
    \includegraphics[width=0.95\textwidth]{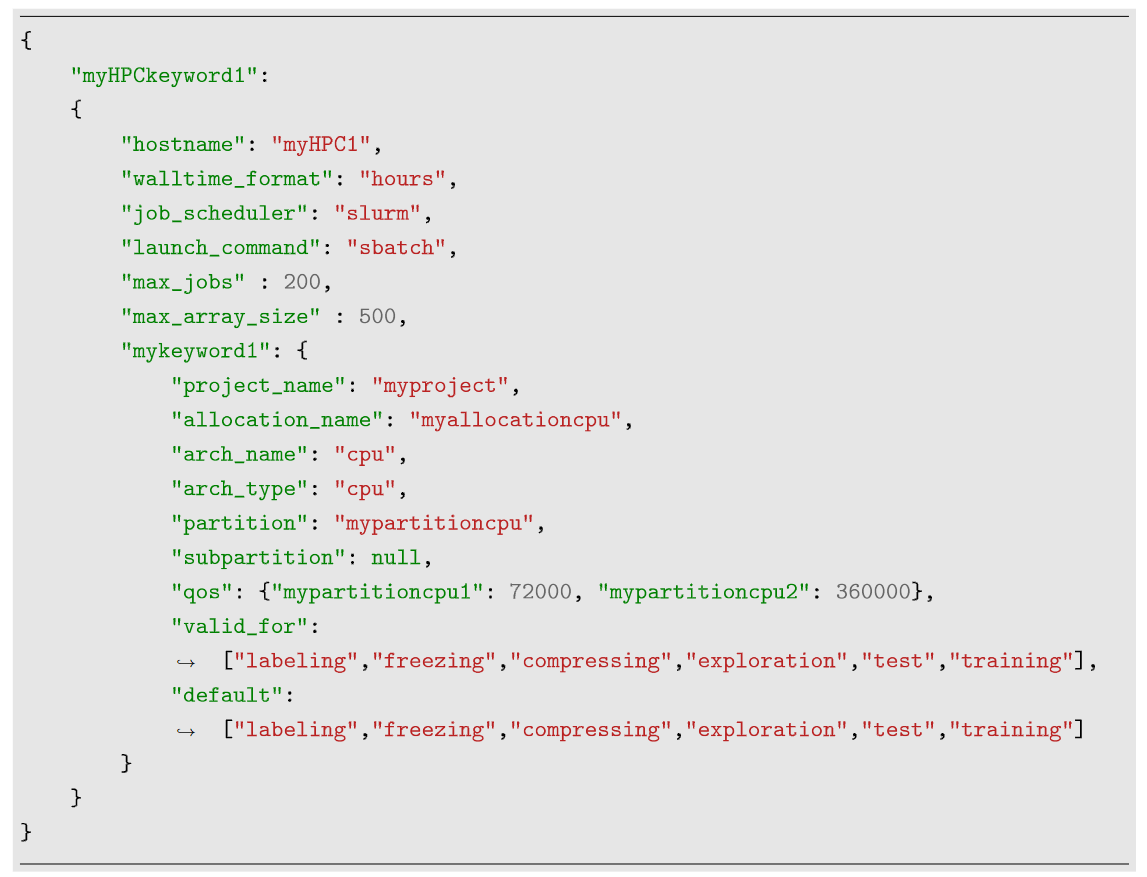}
    \caption{
        Example of a \textit{machine.json} file for configuring HPC resources in ArcaNN.
        }
    \label{fig_si:JSON_machine}
\end{figure}
\clearpage

\subsection{JSON control files written by ArcaNN}

\begin{figure}[!ht]
    \centering
    \includegraphics[width=0.95\textwidth]{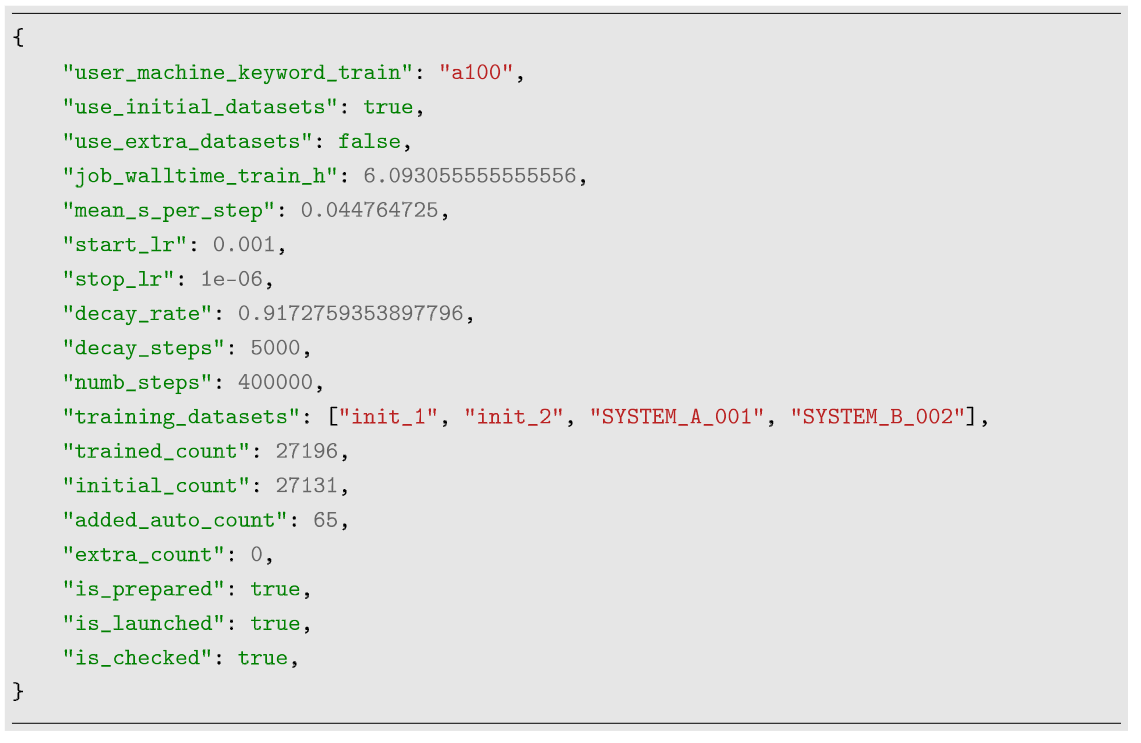}
    \caption{
        A pruned control training JSON file.
        }    \label{fig_si:JSON_training}
\end{figure}

\begin{figure}[!ht]
    \centering
    \includegraphics[width=0.95\textwidth]{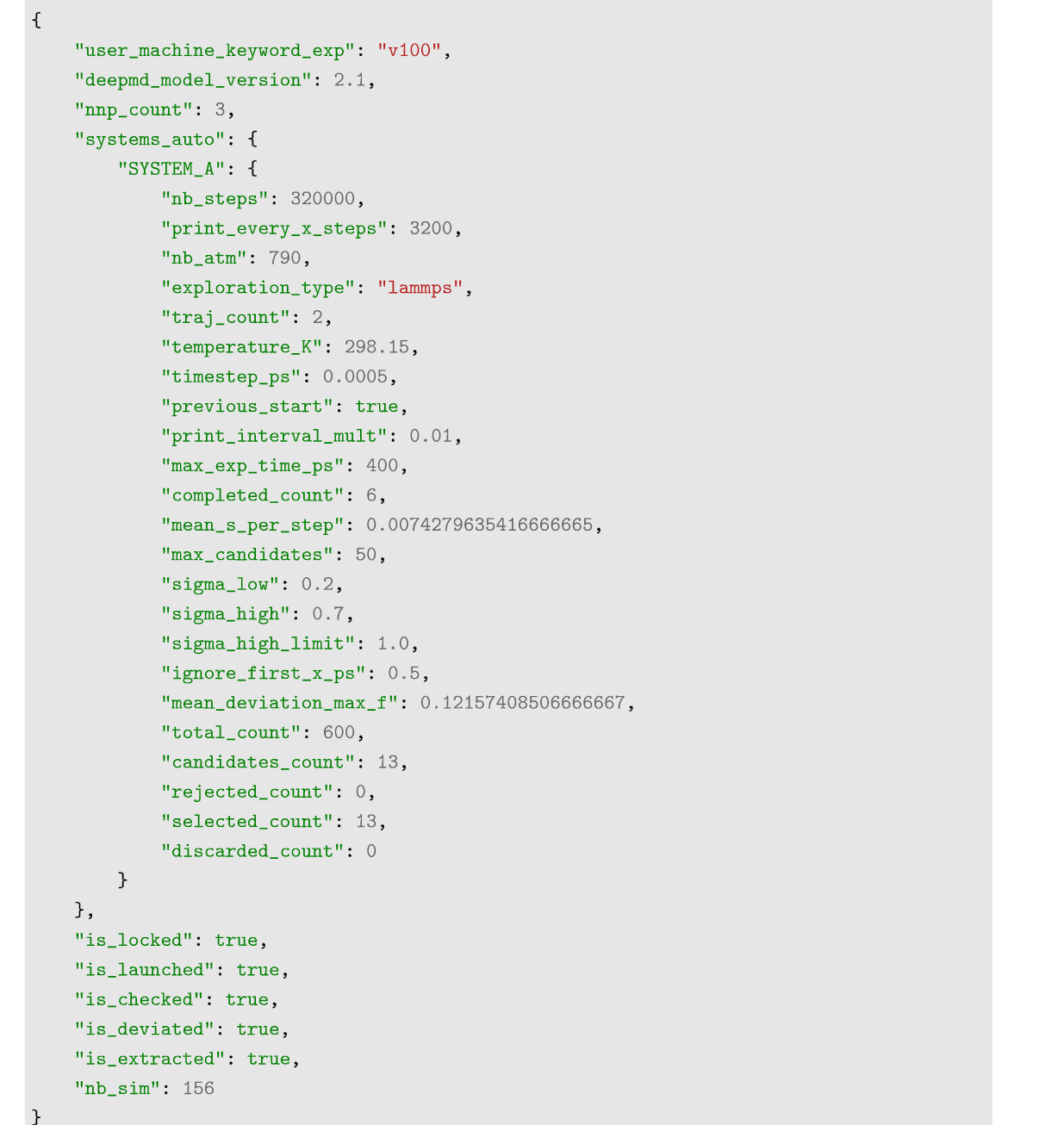}
    \caption{
        A pruned control exploration JSON file.
        }
    \label{fig_si:JSON_exploration}
\end{figure}

\begin{figure}[!ht]
    \centering
    \includegraphics[width=0.95\textwidth]{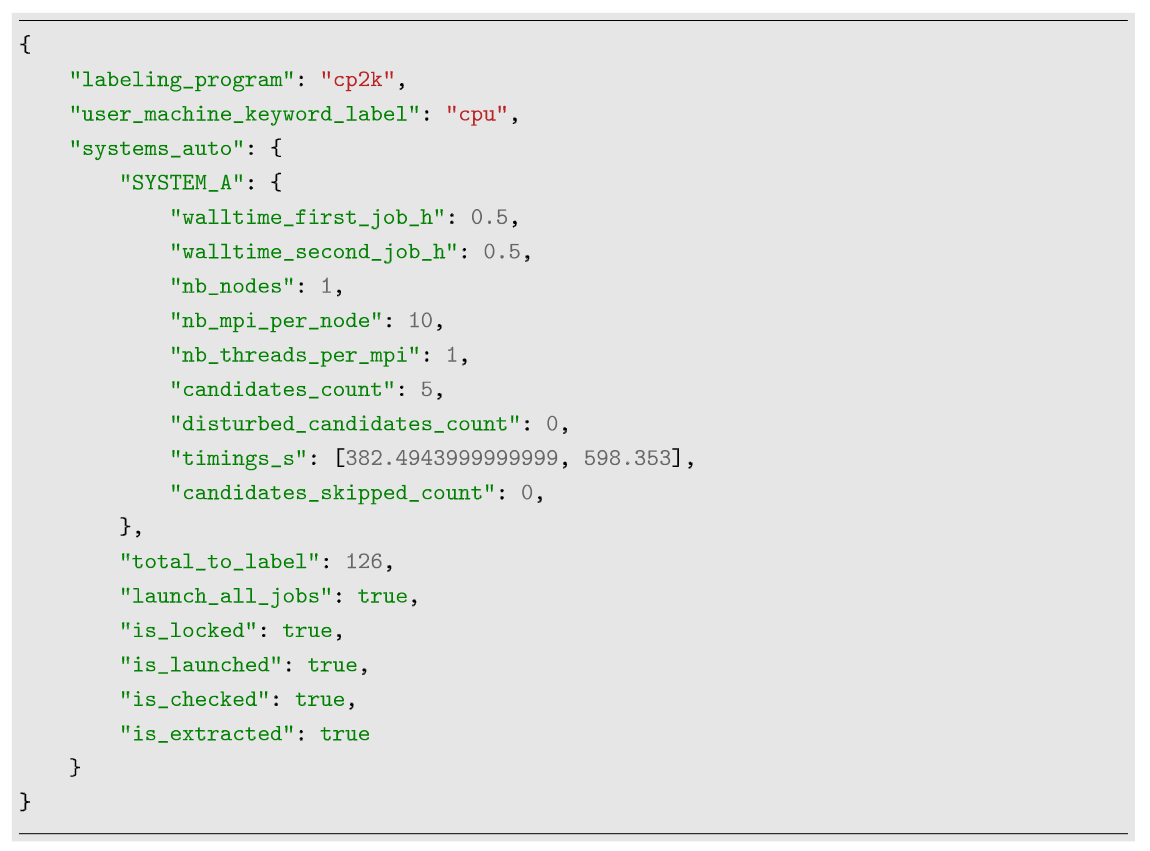}
    \caption{
        A pruned control labeling JSON file.
        }
    \label{fig_si:JSON_labeling}
\end{figure}

\begin{figure}[!ht]
    \centering
    \includegraphics[width=0.95\textwidth]{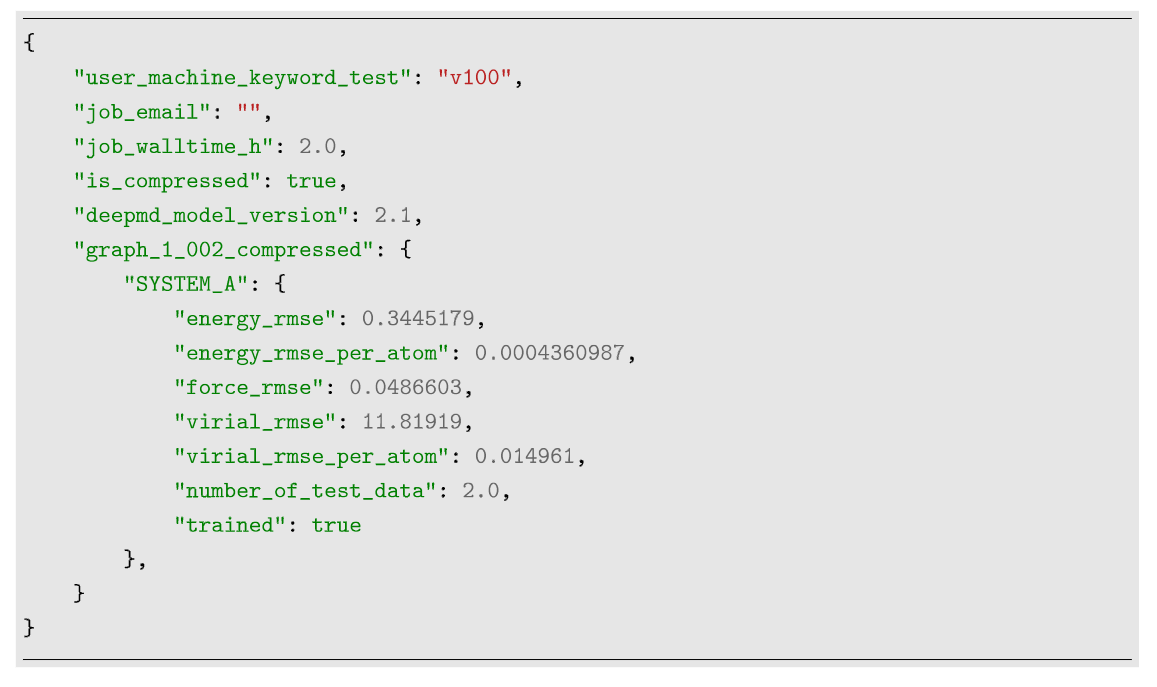}
    \caption{
        A pruned control testing JSON file.
        }
    \label{fig_si:JSON_testing}
\end{figure}
\clearpage

\subsection{Evolution of the candidate and rejected structures with exploration time per reactive exploration iteration for the $S_{N}2$ reaction}

\begin{figure}[!ht]
    \centering
    \includegraphics[width=0.95\textwidth]{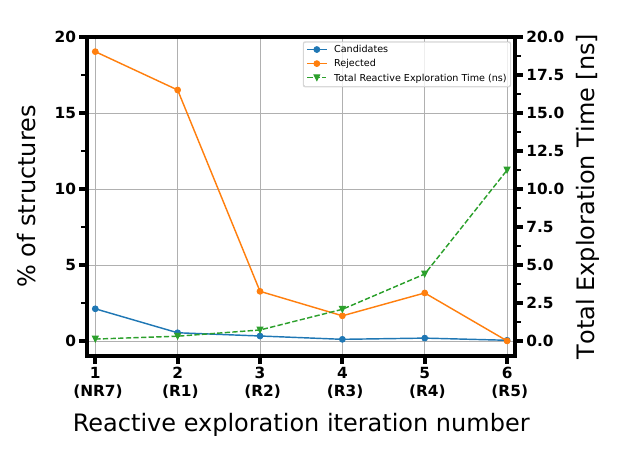}
    \caption{Percentage of candidate structures (solid blue line), rejected structures (solid orange line), and total exploration time (dashed green line) for each reactive exploration step, with the associated training dataset name in parentheses.
    }
    \label{fig_si:r_structures}
\end{figure}
\clearpage

\subsection{Validation of the $R5$ (production) NNP for the $S_{N}2$ reaction}

In Figure \ref{fig_si:rmse_along_pathway}, we report the component-wise force RMSE and the maximum component-wise force error along $\delta d$ on the test dataset.

\begin{figure}[!ht]
    \centering
    \includegraphics[width=0.95\textwidth]{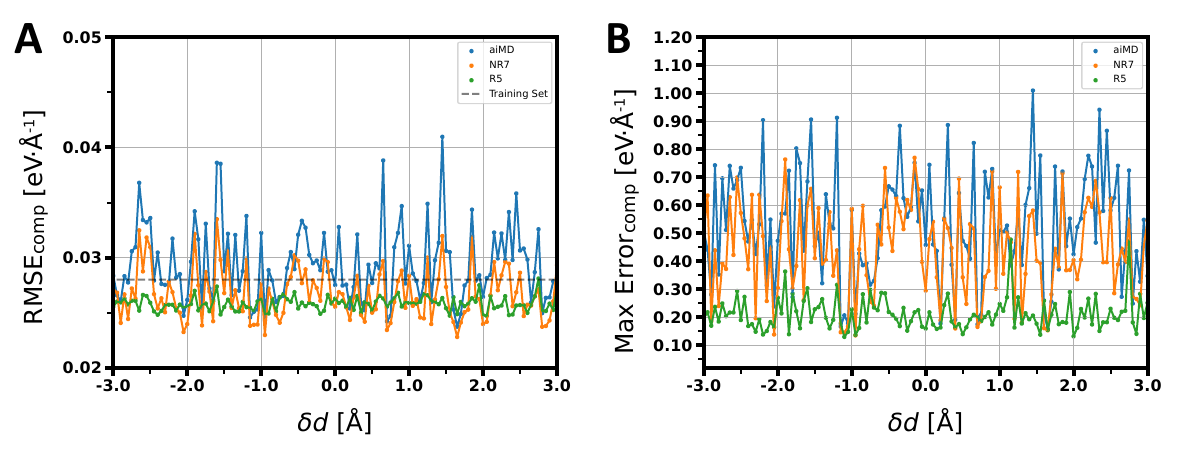}
    \caption{
        For three generation of NNPs, $aiMD$ (blue), $NR7$ (orange) and $R5$ (green):
        (A) the component-wise force RMSE along $\delta d$ on the test dataset.
        (B) the maximum component-wise force error along $\delta d$ on the test dataset.
    }
    \label{fig_si:rmse_along_pathway}
\end{figure}

In Figure \ref{fig_si:errors_on_forces_training_test}, we report the probability density of the component-wise force errors and the probability density of the magnitude per atom force errors on the training and test datasets for the last ($R5$) reactive cycle.
The RMSE on component-wise forces for training dataset is equal to \SI{0.028}{\eV\per\angstrom} and for test dataset is equal to \SI{0.026}{\eV\per\angstrom}, while the RMSE on the magnitude per atom force errors for training dataset is equal to \SI{0.048}{\eV\per\angstrom} and for test dataset is equal to \SI{0.045}{\eV\per\angstrom}.

\begin{figure}[!ht]
    \centering
    \includegraphics[width=0.95\textwidth]{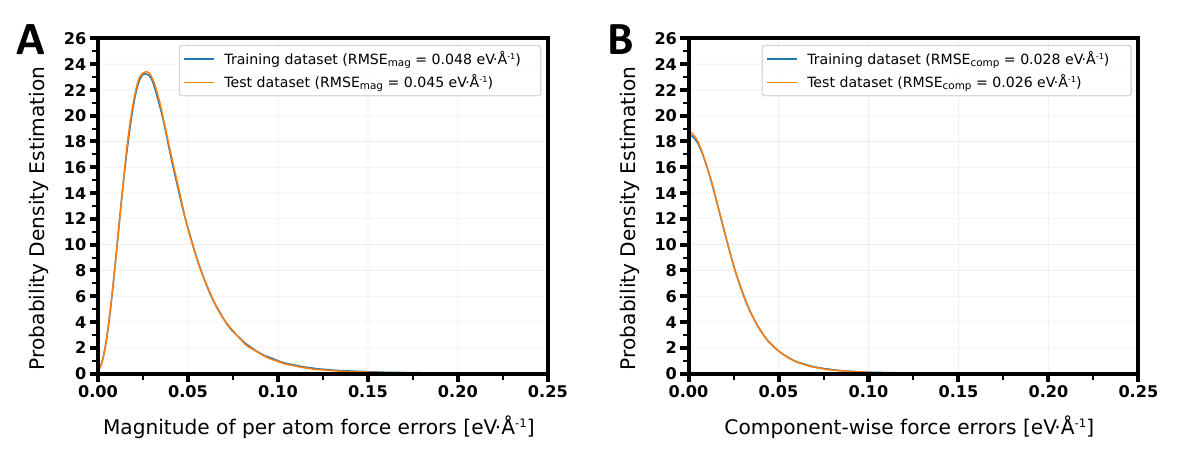}
    \caption{
        (A) Probability density of the magnitude of per-atom force errors on the training dataset and the test dataset.
        (B) Probability density of the component-wise force errors on the training dataset and the test dataset.
    }
    \label{fig_si:errors_on_forces_training_test}
\end{figure}
\clearpage

\subsection{Free energy profiles and CV for the $NR2$ and $NR3$ NNPs for the $S_{N}2$ reaction}

\begin{figure}[!ht]
    \centering
    \includegraphics[width=0.95\textwidth]{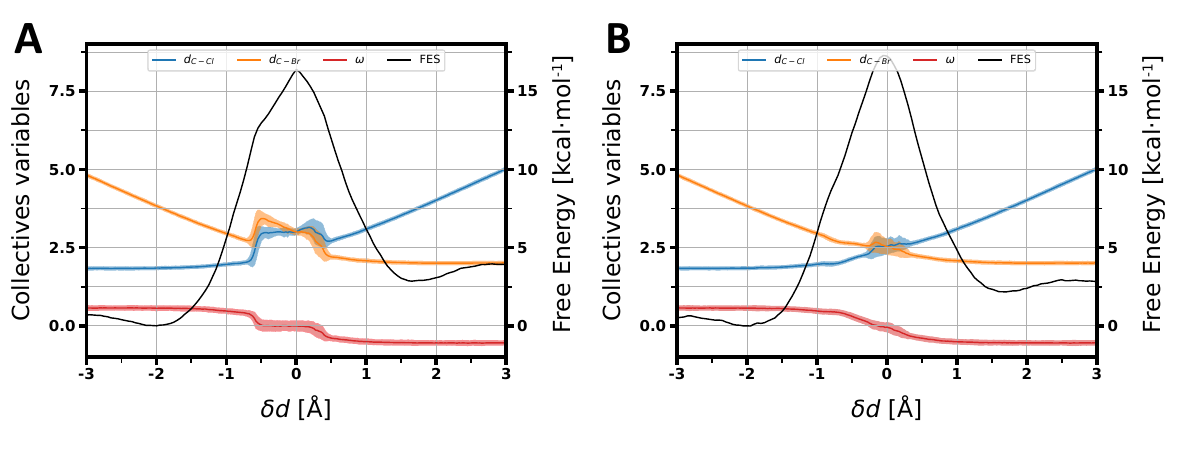}
    \caption{
        Free energy surface obtained from the Umbrella Sampling simulations (black) and the average value of the collective variables (as well as the 95\% confidence interval in shaded color);
        (A) with the NNP trained on the $NR2$ dataset.
        (B) with the NNP trained on the $NR3$ dataset.
    }
    \label{fig_si:fep_cv_us_nr2_nr3}
\end{figure}
\clearpage

\subsection{Validation of the $R8$ (production) NNP for the Diels-Alder reaction}

In Figure \ref{fig_si:errors_on_forces_training_test_da}, we report the probability density of the magnitude per atom force errors on the training and test datasets for the last ($R8$) reactive cycle.
The RMSE on component-wise forces for training dataset is equal to \SI{0.070}{\eV\per\angstrom} and for test dataset is equal to \SI{0.071}{\eV\per\angstrom} (see Figure \ref{fig_main:da_reaction_cv_error_fep}C), while the RMSE on the magnitude per atom force errors for training dataset is equal to \SI{0.122}{\eV\per\angstrom} and for test dataset is equal to \SI{0.123}{\eV\per\angstrom}.

\begin{figure}[!ht]
    \centering
    \includegraphics[width=0.95\textwidth]{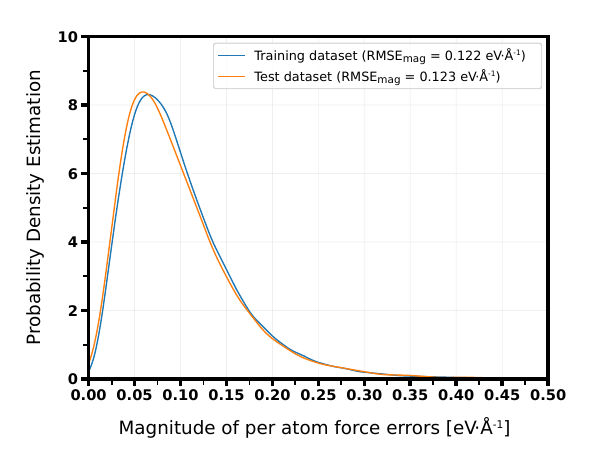}
    \caption{Probability density of the magnitude of per-atom force errors on the training dataset and the test dataset.
    }
    \label{fig_si:errors_on_forces_training_test_da}
\end{figure}

In Figure \ref{fig_si:rmse_along_pathway_da}, we report the component-wise force RMSE and the maximum component-wise force error along $\bar d$ on the test dataset.

\begin{figure}[!ht]
    \centering
    \includegraphics[width=0.95\textwidth]{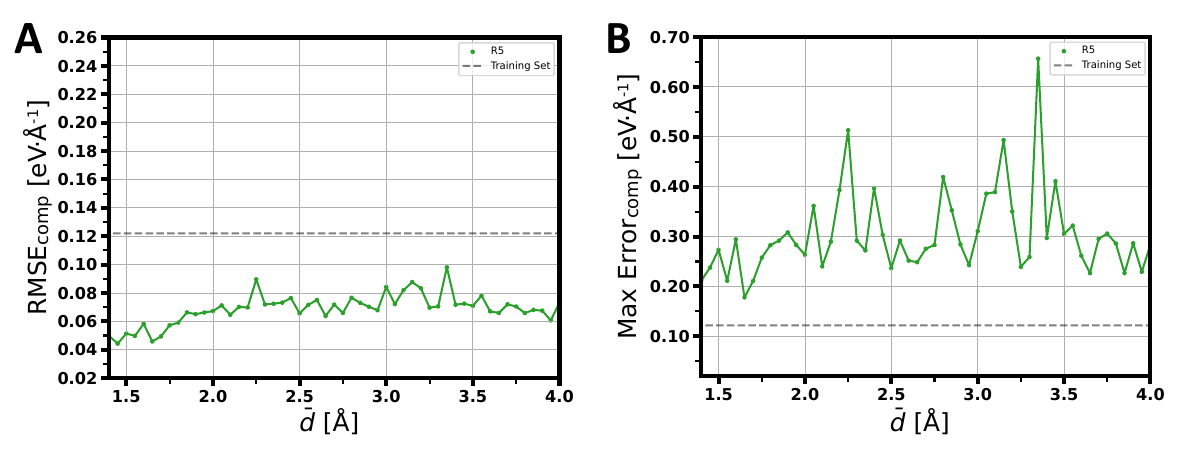}
    \caption{
        (A) the component-wise force RMSE along $\delta d$ on the test dataset for the $R8$ NNP.
        (B) the maximum component-wise force error along $\delta d$ on the test dataset for the $R8$ NNP.
    }
    \label{fig_si:rmse_along_pathway_da}
\end{figure}
\clearpage

\subsection{Joint density distribution of the two main distances for the Diels-Alder reaction}

\begin{figure}[!ht]
    \centering
    \includegraphics[width=0.95\textwidth]{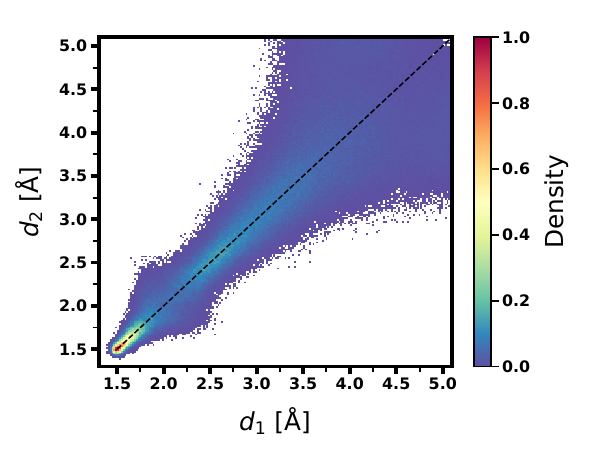}
    \caption{
        Joint density distribution of the distance $d_{1}$ and the distance $d_{2}$, with the dotted line representing $d_{1}$ = $d_{2}$, in the Umbrella Sampling simulations.
    }
    \label{fig_si:distribution_d1_d2_US_da}
\end{figure}
\clearpage

\subsection{Joint density distribution of the two main distances in each important training dataset for the $S_{N}2$ reaction}

\begin{figure}[!ht]
    \centering
    \includegraphics[width=0.95\textwidth]{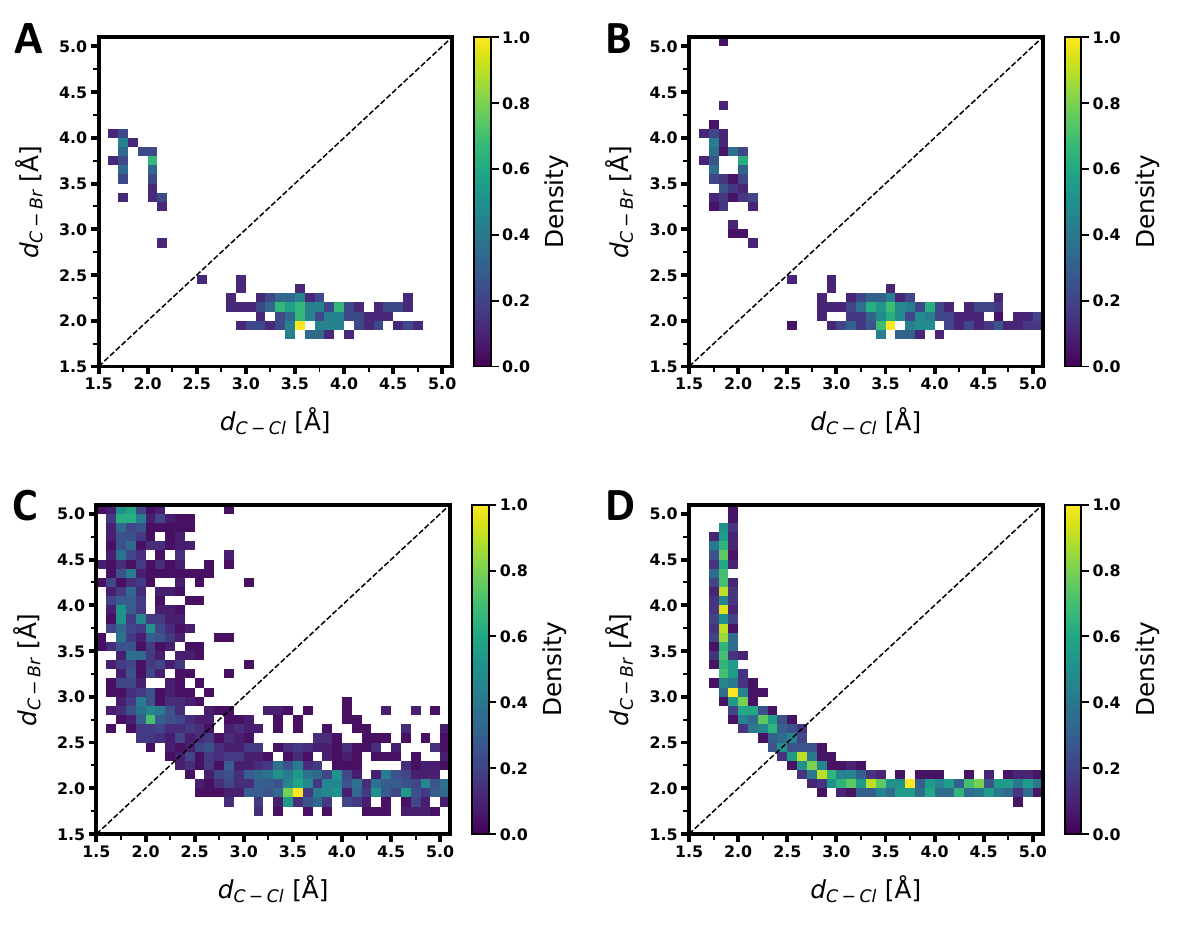}
    \caption{
        Joint density distribution of the distance $d_{C-Cl}$ and the distance $d_{C-Br}$, with the dotted line representing $\delta d$ = \SI{0}{\angstrom} of the structures;
        (A) in the $aiMD$ dataset.
        (B) in the $NR7$ dataset.
        (C) in the $R5$ dataset.
        (D) in the test dataset.
    }
    \label{fig_si:distribution_d1_d2_datasets}
\end{figure}
\clearpage

\subsection{OPES 1D free energy profile and CV from the $R5$ NNP for the $S_{N}2$ reaction}

\begin{figure}[!ht]
    \centering
    \includegraphics[width=0.95\textwidth]{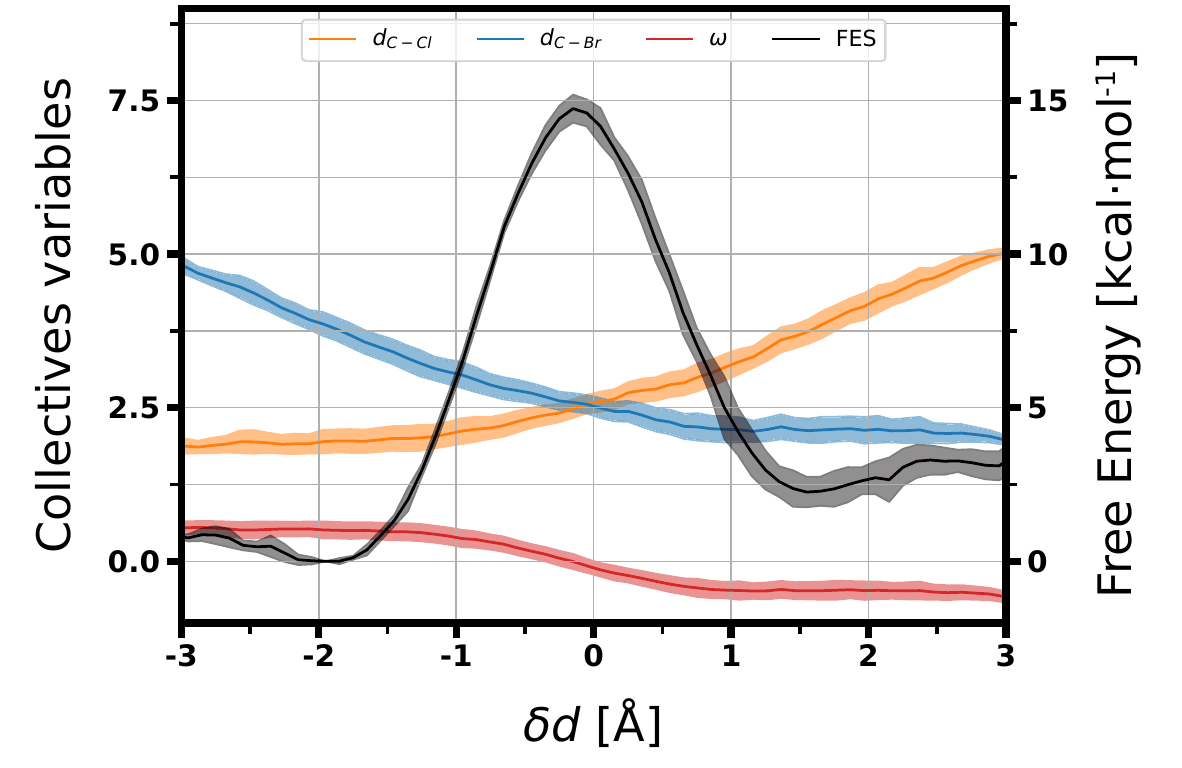}
    \caption{
        Free energy surface obtained from the OPES simulation (with the NNP trainined on the $R5$ dataset) (black) and the average value of the collective variables (as well as the 95\% confidence interval in shaded color)
    }
    \label{fig_si:opes_1d}
\end{figure}
\clearpage

\end{suppinfo}

\bibliography{main}

\end{document}